\documentclass[fleqn,usenatbib]{mnras}

\usepackage{newtxtext,newtxmath}

\usepackage[T1]{fontenc}
\usepackage{ae,aecompl}

\usepackage{times} 
\usepackage{tikz}
\usetikzlibrary{positioning}
\usepackage{xspace,ulem}

\usetikzlibrary{backgrounds}
\usetikzlibrary{arrows}
\usetikzlibrary{calc}
\usepackage{booktabs}
\usepackage{graphicx}	
\usepackage{amsmath}	
\usepackage{amssymb}	
\usepackage{times}

\title[Halo concentration modelling in $f(R)$ gravity]{A general framework to test gravity using galaxy clusters II: \\ A universal model for the halo concentration in $f(R)$ gravity}

\author[M. A. Mitchell et al.]{
Myles A. Mitchell,$^{1}$\thanks{E-mail: m.a.mitchell@durham.ac.uk}
Christian Arnold,$^{1}$
Jian-hua He$^{2,1}$
and Baojiu Li$^{1}$
\\
$^{1}$Institute for Computational Cosmology, Department of Physics, Durham University, South Road, Durham DH1 3LE, UK\\
$^{2}${School of Astronomy and Space Science, Nanjing University, Nanjing 210093, P. R. China}
}

\date{Accepted XXX. Received YYY; in original form ZZZ}

\pubyear{2019}

\begin{document}
\label{firstpage}
\pagerange{\pageref{firstpage}--\pageref{lastpage}}
\maketitle

\begin{abstract}
We present a novel fitting formula for the halo concentration enhancement in chameleon $f(R)$ gravity relative to General Relativity (GR). The formula is derived by employing a large set of $N$-body simulations of the Hu-Sawicki $f(R)$ model which cover a wide range of model and cosmological parameters, resolutions and simulation box sizes. The complicated dependence of the concentration on halo mass $M$, redshift $z$, and the $f(R)$ and cosmological parameters can be combined into a simpler form that depends only on a rescaled mass $M/10^{p_2}$, with $p_2\equiv1.5\log_{10}\left[|{\bar{f}_R(z)}|/(1+z)\right]+21.64$ and $\bar{f}_R(z)$ the background scalar field at $z$, irrespective of the $f(R)$ model parameter. Our fitting formula can describe the concentration enhancement well for redshifts $z\leq3$, nearly 7 orders of magnitude in $M/10^{p_2}$ and five decades in halo mass. This is part of a series of works which aims to provide a general framework for self-consistent and unbiased tests of gravity using present and upcoming galaxy cluster surveys. The fitting formula, which is the first quantitative model for the concentration enhancement due to chameleon type modified gravity, is an important part in this framework and will allow continuous exploration of the parameter space. It can also be used to model other statistics such as the matter power spectrum. 
\end{abstract}




\section{Introduction}
\label{sec:introduction}

Galaxy clusters are the largest-known gravitationally-bound objects in the Universe and are believed to trace the highest peaks of the density perturbations that originated in the early Universe. As a result, the global properties of clusters such as their abundance are highly sensitive to the parameters of cosmological models and theories of gravitation. They are therefore powerful probes of modified gravity theories that lead to different histories of large-scale structure formation than predicted by the standard $\Lambda$CDM paradigm which is based on Einstein's GR.

Various ongoing and upcoming large galaxy and cluster surveys \citep[e.g.,][]{ukidss, desi, euclid, lsst, xmm-newton, chandra, erosita, Planck_SZ_cluster, act} are providing a wealth of information that is likely to revolutionise our constraints of these theories. However in order to make the best possible constraints, it is necessary to prepare a robust theoretical apparatus that can be combined with this data without inducing systematic bias. As an example, tests that use the cluster abundance require a calibration of the halo mass function (HMF) from $N$-body simulations that have been run for the model being constrained. Ideally the calibration will have an explicit dependence on the model parameters and should be in a form in which it may be safely combined with the observational data to constrain these parameters. For example, the definition of the halo mass which is used in the theoretical predictions should be consistent with the method used in the observational survey to measure the cluster mass.

\begin{figure*}
\centering
\begin{tikzpicture}
\tikzstyle{myarrow}=[line width=0.5mm,draw=black,-triangle 45,postaction={draw, line width=0.5mm, shorten >=4mm, -}]

\node    (simulations)    {MG simulation data};
\node    (cat_true)    [below left = 0.5cm and -0.5cm of simulations]   [align=center]{halo catalogue \\ ($M_{\rm true}$)};
\node    (c_m)    [below = 0.75cm of cat_true] [align=center]   {$c_{\rm 200}(M_{\rm 500})$};
\node   (m300_m500)    [below = 1.25cm of c_m] [align=center]    {$\frac{M_{\rm 300m}}{M_{500}}$};
\node    (nfw)    [below left = 0.65cm and 1.0cm of c_m]  [anchor=west]   {NFW};
\node    (hmf_m300)    [below left = 2.7cm and 2.0cm of c_m]  [anchor=west]   {$\left( \frac{{\rm d} n_{\rm halo}}{{\rm d}M_{\rm 300m}} \right)_{f(R)}$};
\node    (hmf_th)    [below = 1.7cm of m300_m500]  [align=center]   {$\left( \frac{{\rm d} n_{\rm halo}}{{\rm d} M_{500}} \right)_{f(R)}$};
\node    (rho_eff)    [below right = 0.5cm and -0.5cm of simulations]   [align=center]  {effective density};
\node    (cat_mdyn)    [below = 1.0cm of rho_eff]    [align=center] [align=center]{halo catalogue \\ ($M_{\rm dyn}$)};
\node    (mdyn_mtrue)    [below = 1.0cm of cat_mdyn]  [align=center]   {$\frac{M_{\rm dyn}}{M_{\rm true}}(M_{\rm true})$};
\node    (observations)    [right = 5cm of simulations]    [align=center] {observational data};
\node    (n_Y)    [below = 2.0cm of observations]   [align=center]  {$\frac{{\rm d} n_{\rm cluster}}{{\rm d}Y_{\rm{obs}}}$};
\node    (hmf_obs)  at (hmf_th -| n_Y)  [ align=center]   {$\left(\frac{{\rm d} n_{\rm cluster}}{{\rm d} M_{\rm 500}}\right)_{f(R)}$};
\node    (scaling_relation)    [right = 1.75cm of mdyn_mtrue]    [align=center]{$Y_{\rm{obs}}^{f(R)}(M_{500})$};
\node    (scaling_relation_lcdm)    [below left= 0.5cm and -0cm of scaling_relation]    [align=center]{$Y_{\rm{obs}}^{\Lambda\rm{CDM}}(M_{500})$};

\node    (mcmc)    at ($(hmf_th)!0.5!(hmf_obs)+(0.0,0.25)$)   [align=center]  {MCMC};
\node    (constraint)  at ($(hmf_th)!0.5!(hmf_obs)+(0.0,-1.0)$) [align=center]  {$|f_{R0}|$ constraint};

\draw[->, line width=0.5mm] (simulations) -- (cat_true);
\draw[->, line width=0.5mm] (cat_true) -- (c_m);
\draw[->, line width=0.5mm] (c_m) -- (m300_m500);
\draw[->, line width=0.5mm, to path={-| (\tikztotarget)}] (nfw) edge (m300_m500);
\draw[->, line width=0.5mm] (m300_m500) -- (hmf_th);
\draw[->, line width=0.5mm, to path={-| (\tikztotarget)}] (hmf_m300) edge (hmf_th);
\draw[->, line width=0.5mm] (simulations) -- (rho_eff);
\draw[->, line width=0.5mm] (rho_eff) -- (cat_mdyn);
\draw[->, line width=0.5mm] (cat_mdyn) -- (mdyn_mtrue);

\draw[->, line width=0.5mm] (observations) -- (n_Y);
\draw[->, line width=0.5mm] (n_Y) -- (hmf_obs);
\draw[->, line width=0.5mm, to path={-| (\tikztotarget)}] (scaling_relation) edge (hmf_obs);
\draw[->, line width=0.5mm] (mdyn_mtrue) -- (scaling_relation);
\draw[->, line width=0.5mm, to path={|- (\tikztotarget)}] (scaling_relation_lcdm) edge (scaling_relation);

\draw[->, line width=0.5mm, to path={-| (\tikztotarget)}] (hmf_th) edge (constraint);
\draw[->, line width=0.5mm, to path={-| (\tikztotarget)}] (hmf_obs) edge (constraint);


\draw [line width=0.5mm,dotted, red, rounded corners=15pt]     ($(rho_eff.north west)+(-0.4,0.15)$) rectangle ($(mdyn_mtrue.south east)+(0.45,-0.1)$);
\node [above right = 0.10cm and -1.3cm of rho_eff] {\small{\color{red}\hypersetup{citecolor=red}\cite{Mitchell:2018qrg}\hypersetup{citecolor=blue}}}; 
\draw[line width=0.5mm,dotted, blue, rounded corners=15pt]   ($(cat_true.north west)+(-0.4,0.15)$) rectangle ($(c_m.south east)+(0.4,-0.1)$);
\node [above left = 0.10cm and -1.3cm of cat_true] {\small{\color{blue}this paper}}; 
\draw[line width=0.5mm,dotted, brown, rounded corners=15pt]     ($(hmf_th.north west)+(-0.4,0.2)$) rectangle ($(constraint.south east -| hmf_obs.south east) +(0.45,-0.1)$);
\node [above right = 0.20cm and -0.9cm of hmf_obs] {\small{\color{brown}future work}}; 
\draw[line width=0.5mm,dotted, black!60!green, rounded corners=15pt]     ($(scaling_relation.north west)+(-0.4,0.1)$) rectangle ($(scaling_relation.south east) +(0.4,-0.1)$);
\node [above right = 0.10cm and -1.3cm of scaling_relation] {\small{\color{black!60!green}future work}}; 

\end{tikzpicture}
\caption{Flow chart outlining the layout of the framework proposed by \citet{Mitchell:2018qrg}. This has been updated from the original version, and illustrates the key steps of the framework to constrain $f(R)$ gravity using the galaxy cluster abundance. Our model for the concentration in $f(R)$ gravity (blue dotted box), which is discussed in this work, can be used to evaluate conversions between different halo mass overdensities. This can be used to ensure that fitting formulae for the halo mass function in $f(R)$ gravity \citep[e.g.,][]{Cataneo_et_al._(2016)} are evaluated with the same mass overdensity as measurements of the mass function from observational surveys. \citet{Mitchell:2018qrg} calibrated a general model for the enhancement of the dynamical mass of haloes in $f(R)$ gravity (red dotted box). This can be used to convert observable-mass scaling relations, which have been calibrated in $\Lambda$CDM, into the correct form in $f(R)$ gravity, using the predictions of \citet{He:2015mva}. Full-physics hydrodynamical simulations will be used in future work (green dotted box) to further test the effect of galaxy formation on the accuracy of these predictions. The corrected scaling relations can be used to infer the mass function from observations. This can be confronted with theoretical predictions to constrain $f(R)$ gravity using Markov chain Monte Carlo methods (brown dotted box).}
\label{fig:flow_chart}
\end{figure*}
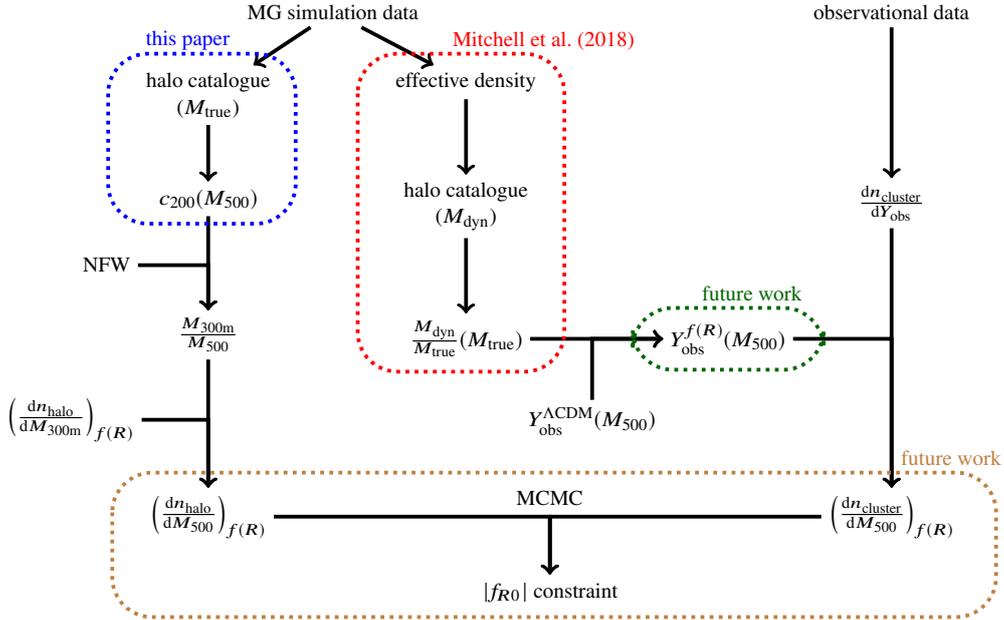

A self-consistent framework, which is introduced in \cite{Mitchell:2018qrg}, is designed to make unbiased constraints of modified gravity theories using galaxy clusters. Our current focus is on using the cluster abundance to constrain the present-day background scalar field of Hu-Sawicki (HS) $f(R)$ gravity \citep{Hu-Sawicki}, which is a representative example of a large class of scalar-tensor gravity theories that can pass stringent solar system tests. The framework is summarised in Fig.~\ref{fig:flow_chart}. This addresses a number of potential  sources of bias. For example, in HS $f(R)$ gravity the total gravitational force that is felt by a massive test particle can be up to a factor $4/3$ stronger than the Newtonian gravitational force. This means that the dynamical mass of an object can be enhanced by up to a factor $4/3$ compared to the true mass \citep{Schmidt:2010jr,Zhao:2011cu,arnold:2014,Gronke:2015,Gronke:2016}. Therefore a model for the HMF that is calibrated using the true mass of haloes cannot be directly combined with observational data that is calibrated using the dynamical mass. 

In \cite{Mitchell:2018qrg} a general analytical model for the enhancement of the dynamical mass (red dotted box of Fig.~\ref{fig:flow_chart}) was calibrated, which depends on the true cluster mass, redshift and the scalar field. This improves on previous studies from the literature which tend to look for a more qualitative understanding or only focus on a particular redshift for a single $f(R)$ gravity model \citep[e.g.,][]{Schmidt:2010jr,Zhao:2011cu,arnold:2014}. The model can be used to predict the dynamical masses of dark matter haloes in mock catalogues and to convert cluster observable-mass scaling relations that have been calibrated in $\Lambda$-cold-dark-matter ($\Lambda$CDM) into a form that works in $f(R)$ gravity \citep{He:2015mva}. Another potential application of the model is in tests of $f(R)$ gravity in which the dynamical mass of objects is compared to their lensing mass. This could, for example, be achieved by comparing the lensing profiles with the X-ray profiles \citep{Terukina:2013eqa,Wilcox:2015kna,Wilcox:2016guw,Pizzuti:2017diz}.

Another important aspect of our framework is the ability to make conversions between halo masses corresponding to different overdensities, which in this work are denoted by the symbol $\Delta$. Note that throughout this paper we define the halo mass $M_{\Delta}$ as the total mass contained within a sphere, of radius $R_{\Delta}$, that encloses an average density of $\Delta$ times the critical density of the Universe. For example, in order to constrain $f(R)$ gravity using the cluster abundance we require a model-dependent calibration of the HMF, which quantifies the number density of dark matter haloes per unit mass interval, ${\rm d}n_{\rm halo}/{\rm d}M_{\Delta}$. Our choice of HMF \citep{Cataneo_et_al._(2016)} has been calibrated for overdensity $\Delta=300\Omega_{\rm M}$, while overdensity $\Delta=500$ is more generally used in cluster surveys. Therefore in order to make constraints using observational data that has been calibrated for overdensity $\Delta=500$, it will be necessary for us to apply the conversion $M_{\rm 300m} \rightarrow M_{500}$ to the HMF. Note that the `m' in the subscript here means that the overdensity is multiplied by $\Omega_{\rm M}$. It is also likely that conversions to other overdensities will be required. For example $\Delta=2500$ is also sometimes used in observational surveys, and $\Delta=200$ is often used in theoretical studies. Therefore a prediction for the conversion between halo masses corresponding to arbitrary values of $\Delta$ is essential. 

The halo mass measured for different overdensities corresponds to the total mass enclosed by different halo radii $R_{\Delta}$, where $R_{\Delta}$ is larger for lower values of $\Delta$. Therefore conversions between the halo mass at different overdensities can be estimated if the density profile of dark matter haloes can be predicted. Typically the universal Navarro-Frenk-White (NFW) density profile \citep{NFW} is assumed. This is a 2-parameter profile, but can be written with one parameter if the mass (or radius) for a particular overdensity is known. This parameter can be the concentration, and predicting it in $\Lambda$CDM as a function of the cluster mass and redshift has been the subject of much work over the two decades since it was first introduced \citep[e.g.,][]{Bullock:1999he,Neto:2007vq,Duffy:2008pz,Dutton:2014xda,Ludlow:2013vxa}.

In addition to facilitating mass conversions the concentration is also important in studies of the non-linear matter power spectrum \citep[e.g.,][]{Brax:2013fna,Lombriser:2013eza,2016PhRvD..93j3522A,Hu:2017aei,2018arXiv181205594C}, which, like the cluster abundance, can also be used to probe dark energy and modified gravity theories. The large-scale part of the matter power spectrum can, for example, be predicted using linear perturbation theory by incorporating the linear halo bias. On the other hand, the small-scale part of the matter power spectrum can be assembled using the HMF and the halo concentration. The concentration is necessary in order to predict the density profile, which is required, for example, in order to model the size of haloes.

In $f(R)$ gravity the concentration can become enhanced due to the effects of the fifth force on the density profile. For example, for an unscreened halo the in-falling particles experience a greater acceleration due to the stronger gravitational force, and this can alter the profile such that the density is raised at the inner regions and lowered at the outer regions. Therefore the $\Lambda$CDM predictions of the concentration are unlikely to apply for lower-mass unscreened haloes in $f(R)$ gravity. Yet there is no general quantitative model for the concentration in $f(R)$ gravity that is discussed in the literature, which instead tends to focus on a more qualitative understanding of the effects of the fifth force on the concentration and on the density profile \citep[e.g.,][]{Zhao:2010qy,Lombriser:2012nn,Shi:2015aya,arnold:2016,Arnold:2018nmv}. Therefore a $\Lambda$CDM relation for the concentration is often used in the literature. For example, the modellings of the non-linear matter power spectrum in HS $f(R)$ gravity by \citet{Brax:2013fna,Hu:2017aei,2018arXiv181205594C} use prescriptions for the concentration-mass relation that have been calibrated in $\Lambda$CDM. Also, due to the large scatter of the concentration-mass relation, some works \citep[e.g.,][]{Cataneo_et_al._(2015)} argue that it is fine to assume a fixed concentration for a sample of clusters that covers a sufficiently narrow mass range. 

In order to prevent potential biases resulting from a simplified treatment of the concentration, the focus of this paper is to produce a general model for the concentration in HS $f(R)$ gravity (blue dotted box of Fig.~\ref{fig:flow_chart}) that may be applied in future studies. Rather than calibrating a relation for the absolute concentration, we decided to focus on finding a universal model for the enhancement of the concentration as a function of the halo mass and redshift. This has been achieved using data from a suite of dark-matter-only $N$-body simulations run for three models of HS $f(R)$ gravity. Note that we define the enhancement as the ratio of the $f(R)$ concentration to the concentration in GR. This means that the reader can select a $\Lambda$CDM concentration-mass-redshift relation from the literature \citep[e.g.,][]{Bullock:1999he,Neto:2007vq,Duffy:2008pz,Dutton:2014xda,Ludlow:2013vxa} that they wish to use, then this can be converted into a form in HS $f(R)$ gravity. Our model includes a dependence on the cosmological density parameters $\Omega_{\rm M}$ and $\Omega_{\Lambda}$, so any $\Lambda$CDM relation can be used regardless of the values of these parameters. Our model depends on the particular combination $\bar{f_R}(z)/(1+z)$ where $\bar{f_R}(z)$ is the background scalar field at redshift $z$, and does not explicitly depend on the model parameter $f_{R0}$, i.e., the present-day background scalar field value, as one would naively expect. This has the implication that predictions may be made for arbitrary values of $f_{R0}$ and $z$ (as long as the above combination is within the range of validity of our fitting). This generality of our model was achieved by combining data from the different $f(R)$ gravity models by applying a simple transformation to the halo mass using the $p_2$ parameter defined by \citet{Mitchell:2018qrg}, where $M_{500}=10^{p_2}h^{-1}M_{\odot}$ can be considered as the mass above which haloes are screened and below which haloes are unscreened.

The paper is arranged as follows: Sec.~\ref{sec:theory} introduces the background theory that is relevant to this work; Sec.~\ref{sec:simulations_and_methods} provides an overview of the dark-matter-only $N$-body simulations that are used in this work, along with an outline of the methods used to measure the concentration and its enhancement; Sec.~\ref{sec:results} discusses the results of this work, including the general model for the concentration enhancement; and, finally, Sec.~\ref{sec:conclusions} summarises the main conclusions from this study, and outlines the next steps of our framework.

Throughout this paper we use the unit convention $c=1$ where $c$ is the speed of light. Greek indices run over $0,1,2,3$ while Roman indices run over $1,2,3$. Unless otherwise stated, an over-bar ($\bar{x}$) denotes the mean background value of a quantity while a subscript `0' means the present-day value.

\section{Theory}
\label{sec:theory}

The theory of $f(R)$ gravity \citep[see, e.g.,][for reviews]{Sotiriou:2008rp,DeFelice:2010aj} can be considered as a modification of the Einstein-Hilbert action of GR, whereby a non-linear scalar function $f(R)$ of the Ricci scalar curvature, $R$, is added to the $R$ term in the integrand. This modified action is given by,
\begin{equation}
S=\int {\rm d}^4x\sqrt{-g}\left[\frac{R+f(R)}{16\pi G}+\mathcal{L}_{\rm M}\right],
\label{eq:action}
\end{equation}
where $g$ is the determinant of the metric tensor, $G$ is the universal gravitational constant and $\mathcal{L}_{\rm M}$ is the Lagrangian density of matter fields. Note that $\Lambda$CDM is restored by taking $f=-2\Lambda$. Taking the variation of Eq.~(\ref{eq:action}) with respect to the metric tensor leads to the modified Einstein field equations: 
\begin{equation}
G_{\alpha \beta} + X_{\alpha \beta} = 8\pi GT_{\alpha \beta},
\label{eq:modified_field_equations}
\end{equation}
where $G_{\alpha \beta}$ is the Einstein tensor and $T_{\alpha \beta}$ is the stress-energy tensor. The modifications to GR are denoted by the term $X_{\alpha \beta}$:
\begin{equation}
X_{\alpha \beta} = f_RR_{\alpha \beta} - \left(\frac{f}{2}-\Box f_R\right)g_{\alpha \beta} - \nabla_{\alpha}\nabla_{\beta}f_R,
\label{eq:GR_modification}
\end{equation}
where $\Box$ is the d'Alembert operator, $R_{\alpha \beta}$ is the Ricci curvature and $\nabla_{\alpha}$ and $\nabla_{\beta}$ denote the covariant derivatives associated with the metric $g_{\alpha\beta}$. Eq.~(\ref{eq:GR_modification}) also includes an extra scalar degree of freedom which is called the scalar field, or "scalaron", and denoted by $f_R\equiv{\rm d}f(R)/{\rm d}R$. This mediates an attractive force, known as the fifth force, whose physical range is set by the Compton wavelength, $\lambda_{\rm C}$, with,
\begin{equation}
\lambda_{\rm C} = a^{-1}\left(3\frac{{\rm d}f_R}{{\rm d}R}\right)^{\frac{1}{2}},
\label{eq:compton_wavelength}
\end{equation}
where $a$ is the cosmic scale factor. On scales smaller than $\lambda_{\rm C}$, if the fifth force is unscreened (see below) then the gravitational forces are enhanced by a factor $4/3$, which enhances the growth of structure \citep{Zhao:2010qy}.

Solar system tests have confirmed GR to a remarkably high precision in our local neighbourhood \citep{Will:2014kxa}. In order to avoid conflict with these tests, the chameleon screening mechanism \citep[e.g.,][]{Khoury:2003aq,Khoury:2003rn,Mota:2006fz} was proposed and used to give the scalar field an environment-dependent effective mass, such that the fifth force is suppressed in deep potential wells. As a result, in dense regions, such as the inner regions of galaxy clusters, the modification of GR is suppressed, while in lower-density regions, such as the outer regions of galaxy clusters, the fifth force is able to act.

It is essential that the functional form of $f(R)$ is chosen so that it gives rise to the late-time cosmic acceleration without violating the solar system constraints \citep[e.g.,][]{Li:2007xn,Brax:2008hh}. One of the most popular of the viable models is the HS model, which was proposed by \cite{Hu-Sawicki}:
\begin{equation}
f(R) = -m^2\frac{c_1\left(-R/m^2\right)^n}{c_2\left(-R/m^2\right)^n+1},
\label{eq:hu_sawicki}
\end{equation}
where $m^2\equiv8\pi G\bar{\rho}_{\rm M,0}/3=H_0^2\Omega_{\rm M}$. In the latter, $\bar{\rho}_{\rm M,0}$ is the mean background matter density, $H_0$ is the Hubble expansion rate today and $\Omega_{\rm M}$ is the matter density parameter. If the background curvature satisfies the inequality $-\bar{R}\gg m^2$, then we have $f(\bar{R})\approx-m^2c_1/c_2$ which is a constant. We can choose $c_1/c_2=6\Omega_{\Lambda}/\Omega_{\rm M}$, where $\Omega_\Lambda\equiv1-\Omega_{\rm M}$, such that,
\begin{equation}
-\bar{R} = 3m^2\left(a^{-3}+4\frac{\Omega_\Lambda}{\Omega_{\rm M}}\right) \approx 3m^2\left(a^{-3}+\frac{2}{3}\frac{c_1}{c_2}\right).
\label{eq:R}
\end{equation}
This indicates that $f(R)$ gravity behaves like a cosmological constant in background cosmology. Note that for a realistic choice of cosmological parameters $-\bar{R}\gg m^2$ is a good approximation.

In the HS model one can simplify the expression for the background scalar field using the approximation $-\bar{R}\gg m^2$:
\begin{equation}
\bar{f_R} = -\frac{c_1}{c_2^2}\frac{n\left(\frac{-\bar{R}}{m^2}\right)^{n-1}}{\left[\left(\frac{-\bar{R}}{m^2}\right)^n+1\right]^2} \approx -n\frac{c_1}{c_2^2}\left(\frac{m^2}{-\bar{R}}\right)^{n+1},
\label{eq:scalar_field}
\end{equation}
in which,
\begin{equation}
\frac{c_1}{c_2^2} = -\frac{1}{n}\left[3\left(1+4\frac{\Omega_{\Lambda}}{\Omega_{\rm M}}\right)\right]^{n+1}f_{R0}.
\label{eq:c1/c22}
\end{equation}
Note that $f_{R0}$ denotes the present-day background value of $f_R$. In this work we omit the over-bar for $f_{R0}$ even though this is a background quantity. Having fixed the value of $c_1/c_2$ in the way described above, the HS model can then be described in terms of just two free model parameters, $n$ and $f_{R0}$. We adopt the values $n=1$ and $|f_{R0}|=10^{-4}, 10^{-5} \text{ or } 10^{-6}$ (F4, F5 or F6, respectively) for all numerical simulations used in this work (see Sec.~\ref{sec:simulations}). Note that the scalar field drops with increasing redshift, and so the effects of $f(R)$ gravity are expected to vanish at high redshift.

In \cite{Mitchell:2018qrg} the thin-shell model \citep{Khoury:2003rn} was used along with a suite of dark-matter-only $N$-body simulations to calibrate a general analytical model for the enhancement of the dynamical mass in $f(R)$ gravity:
\begin{equation}
\frac{M_{\rm dyn}}{M_{\rm true}} = \frac{7}{6}-\frac{1}{6}\tanh\left(p_1\left[\log_{10}\left(M_{\rm true}\right)-p_2\right]\right),
\label{eq:enhancement}
\end{equation}
where $M_{\rm dyn}$ is the dynamical mass and $M_{\rm true}\equiv M_{500}$ is the true mass. Note that, as discussed in Sec.~\ref{sec:introduction}, the overdensity 500 here is with respect to the critical density of the Universe. The reciprocal of the parameter $p_1$ is proportional to the width of the transition of the enhancement between values 1 and $4/3$, and it was found that this is approximately constant: $p_1=(2.21\pm0.01)$. On the other hand, $p_2$ is defined as the logarithm of the halo mass, $\log_{10}(M_{500}M_{\odot}^{-1}h)$, at which the enhancement of the dynamical mass is $7/6$, which is half of the maximum possible enhancement. It was found that $p_2$ varies linearly as a function of the logarithm of a combination of the background scalar field and redshift:
\begin{equation}
p_2=(1.503\pm0.006)\log_{10}\left(\frac{|\bar{f}_R|}{1+z}\right)+(21.64\pm0.03).
\label{eq:p2}
\end{equation}
The dynamical mass is defined as the mass that is felt by a massive test particle assuming Newtonian gravity, and so the enhancement of the dynamical mass is actually equivalent to the enhancement of the total gravitational force (a combination of the Newtonian force and the fifth force) that is felt by the particle. Therefore $M_{500} = 10^{p_2}h^{-1}M_{\odot}$ can be considered as the halo mass above which the halo is screened and below which the halo is unscreened.

\section{Simulations and Methods}
\label{sec:simulations_and_methods}

The simulations that we use in this work are presented in Sec.~\ref{sec:simulations}, along with the methods that we use to extract their halo data. Our methods to measure the concentration and its enhancement and a useful technique of rescaling the halo mass are discussed in Sec.~\ref{sec:methods}.

\subsection{Simulations}
\label{sec:simulations}

Our dark-matter-only simulations have been run using the \textsc{ecosmog} code \citep{Li:2011vk}, which is based on the publicly available $N$-body and hydrodynamical code \textsc{ramses} \citep{Teyssier:2001cp}, and the \textsc{arepo} code \citep{2010MNRAS.401..791S} with its modified gravity solver. These are efficiently parallelised codes that can be used to run cosmological $N$-body simulations for a range of modified gravity scenarios, including $f(R)$ gravity. Both codes use adaptive mesh refinement to ensure accuracy of the fifth force solution in high-density regions. 

\begin{table*}
\centering

\small
\begin{tabular}{ ccccc } 
 \toprule
 
 Parameters and & \multicolumn{4}{c}{Simulations} \\
 models & Diamond & Jade & Crystal & \textsc{arepo} \\

 \midrule

 box size / $h^{-1}$Mpc & 64 & 450 & 1024 & 62 \\ 
 particle number & $512^3$ & $1024^3$ & $1024^3$ & $512^3$ \\ 
 particle mass / $h^{-1}M_{\odot}$ & $1.52\times10^8$ & $6.64\times10^9$ & $7.78\times10^{10}$ & $1.52\times10^8$ \\
 & & & & \\
 $\Omega_{\rm M}$ & 0.281 & 0.2819 & 0.281 & 0.3089 \\ 
 $\Omega_{\Lambda}=1-\Omega_{\rm M}$ & 0.719 & 0.7181 & 0.719 & 0.6911 \\
 $h$ & 0.697 & 0.697 & 0.697 & 0.6774 \\
 $f(R)$ models & F6 & F5 & F4, F5, F6 & F4, F5, F6\\
 
 \bottomrule
 
\end{tabular}

\caption{Specifications of the three \textsc{ecosmog} simulations and the \textsc{arepo} simulation used in this investigation. The \textsc{ecosmog} simulations are labelled Diamond, Jade and Crystal for convenience. All simulations have been run for $\Lambda$CDM in addition to the $f(R)$ gravity models listed, where F4, F5, and F6 correspond to present-day scalar field strengths $|f_{R0}|=10^{-4},10^{-5},10^{-6}$ for Hu-Sawicki $f(R)$ gravity with parameter $n=1$. The Hubble constant, $H_0$, is equal to $100h$ kms$^{-1}$Mpc$^{-1}$ for each simulation.}
\label{table:simulations}

\end{table*}

The cosmological parameters and technical specifications of the simulations used in this investigation are listed in Table~\ref{table:simulations}. 
The three simulations that have been run using the \textsc{ecosmog} code are labelled \textit{Crystal}, \textit{Jade} and \textit{Diamond} in order of increasing resolution. 
The \textsc{arepo} and Diamond simulations both have a mass resolution of $M_{\rm p}=1.52\times10^{8}h^{-1}M_{\odot}$ and resolve lower-mass halos. On the other hand Crystal has a relatively low mass resolution, but its larger box size allows us to investigate high-mass haloes. Jade is able to provide overlapping halo masses with Crystal and the other simulations, allowing haloes with a wide and continuous range of masses to be studied. This is essential in order to comprehensively study the halo concentration across the full transition between the screened and the unscreened regimes. The \textsc{arepo} data is the DM-only subset of the SHYBONE simulation suite \citep{arnold2019} and is particularly useful because it has been run for all three $f(R)$ gravity models examined in this work. Its low particle mass allows low-mass, unscreened haloes to be studied in all three models, ensuring a more detailed exploration of the transition between the screened and unscreened regimes. In addition to this, the similar resolutions of the \textsc{arepo} simulation and Diamond allow a consistency test of the \textsc{ecosmog} and \textsc{arepo} simulations, which is necessary due to the potential disparities between the results from these two codes which employ different algorithms and assume different cosmological parameters.

The simulation data covers redshifts up to at least $z=1$ for all simulations, as the results of this work are initially intended to be used for tests of gravity using the Planck 2015 data \citep{Planck_SZ_cluster}, which covers galaxy clusters with $z<1$. However redshifts $z<2$ and $z<3$ have been included for \textsc{arepo} F5 and F4, respectively, as otherwise the data from these models would only cover the unscreened regime. The dataset consists of 19 snapshots from Crystal, 33 from Jade and 44 from Diamond. For \textsc{arepo} there are 46, 37 and 24 snapshots from F4, F5 and F6, respectively.

The halo catalogues that we construct consist of dark matter haloes identified using the friends-of-friends algorithm (FOF), and include data on the bound substructures within every FOF group which have been identified by the \textsc{subfind} code \citep{subfind}. In this work, the halo mass $M_{\Delta}$ is defined as the total mass enclosed by a sphere of radius $R_{\Delta}$ within which the average density is equal to $\Delta$ times the critical density of the Universe. This sphere is centred at the potential minimum of the halo. Two methods that can be used to calculate the halo concentration (see Sec.~\ref{sec:c_measurement}) also require measurements of the maximum circular velocity, $V_{\rm max}$, and the corresponding orbital radius, $R_{\rm max}$. The \textsc{subfind} code calculates these quantities using just the bound particles of the central, dominant subhalo. To a good approximation, these measurements can be used to represent the maximum circular velocity and the corresponding orbital radius for the entire FOF group. We have checked that within $R_{\rm max}$ almost all particles are bound regardless of whether the fifth force is felt or not.

In order to accurately measure the halo concentration, it is important that the halo consists of enough particles so that it is well-resolved at both the inner and outer regions. Therefore in all of our analyses in this work we only use haloes with more than 1000 particles contained within $R_{500}$. This corresponds to minimum halo masses of $M_{500}=(1.52\times10^{11}$,$6.64\times10^{12}$,$7.78\times10^{13}$,$1.52\times10^{11})h^{-1}M_{\odot}$ for Diamond, Jade, Crystal and \textsc{arepo} respectively.

We bin our haloes by $M_{500}$ (see Sec.~\ref{sec:c_enhancement_measurement}), and our model for the enhancement of the concentration is designed to predict the concentration in $f(R)$ gravity as a function of $M_{500}$ (see Sec.~\ref{sec:general_model}). The reason for choosing overdensity $\Delta=500$ is to be consistent with \cite{Mitchell:2018qrg}, in which $M_{500}$ was used to study the enhancement of the dynamical mass and, crucially, to define the parameter $p_2$ (see Sec.~\ref{sec:theory}). As will be discussed later (Sec.~\ref{sec:rescaled_mass}), the halo mass can be rescaled by this parameter in order to combine data from snapshots with different values of $|\bar{f}_R|/(1+z)$.

Various works in the literature which study the halo evolution and aim to model the concentration as a function of redshift and mass often use relaxed samples \citep[e.g.,][]{Neto:2007vq}. Haloes which have undergone recent mergers are unlikely to give reliable estimates of the concentration. However, we have decided to include all haloes (that satisfy the above mass criteria) since one of the applications of our results will include matter power spectrum predictions, which requires the use of all haloes.

\subsection{Methods}
\label{sec:methods}

In this section we present the three approaches for measuring the halo concentration that are used in this work (Sec.~\ref{sec:c_measurement}), a useful rescaling of the halo mass (Sec.~\ref{sec:rescaled_mass}) and our method of binning the concentration and evaluating its enhancement (Sec.~\ref{sec:c_enhancement_measurement}).

\subsubsection{Concentration measurement}
\label{sec:c_measurement}

Three methods were considered for the measurement of the halo concentration. The most accurate is to directly fit the profiles of the haloes using the NFW profile \citep{NFW}:
\begin{equation}
\rho(r) = \frac{\rho_{\rm s}}{(r/R_{\rm s})(1+r/R_{\rm s})^2},
\label{eq:nfw}
\end{equation}
where $\rho_{\rm s}$ is the characteristic density and $R_{\rm s}$ is the scale radius. The concentration is defined as $c_{200}=R_{200}/R_{\rm s}$. Note that the convention to define the concentration with respect to overdensity 200 is frequently used by the literature. Therefore we elected to use this definition, even though our haloes are required to be binned by $M_{500}$ (see above). This approach is still consistent because as long as the concentration can be predicted for one overdensity, it can also be predicted for overdensity 500.

Radial bins that are equally spaced in logarithmic distance from the halo centre were used to ensure that both the inner and outer regions were equally well-fitted. As shown by \cite{Neto:2007vq}, the choice of radial range over which the profile is fitted can be important. Resolution effects can occur at the outer less-dense regions of the halo or the innermost regions where, due to the limited number of particles, the density can be underestimated. In order to avoid these effects we chose to calculate the densities of 20 logarithmic radial bins, spanning distances $0.05R_{200}$ to $R_{200}$ from the halo centre, which is consistent with the range used by \cite{Neto:2007vq}. These densities were fitted using the formula,
\begin{equation}
\log_{10}(\rho) = \log_{10}(\rho_{\rm s})-\log_{10}(xc_{200})-2\log_{10}(1+xc_{200}),
\label{eq:nfw_fit}
\end{equation}
where $x=r/R_{200}$. This was achieved via unweighted least squares, where $\rho_{\rm s}$ and $c_{200}$ are allowed to vary independently. The halo concentration was set equal to the optimal value of $c_{200}$.

The concentration was originally defined by \cite{NFW} as a parameter of the NFW profile. Therefore fitting this profile to individual haloes is the only means of accurately measuring the concentration in a way that is true to its definition\footnote{We note that for the full NFW fitting one can choose to fit the mass profiles instead of the density profiles of haloes \citep[e.g.,][]{2013ApJ...768..123K}. We consider both cases as full NFW fitting, because they make use of the whole range of halo radius (neglecting certain regions excluded from the fitting).}. Even for an unscreened halo in $f(R)$ gravity, where the density profile can in principle deviate from an NFW profile, the concentration should still be measured in the same way. However there are a number of simplified methods that have been adopted in the literature, including those presented by \cite{Prada:2011jf} (P12) and \cite{Springel:2008cc} (S08), which we also consider in this work. Both methods assume that the halo is well-characterised by the NFW profile without performing a direct fitting. This allows the concentration to be predicted with more limited information, which can save time. 

The P12 method uses the relation between $V_{\rm max}$ and the circular velocity at the halo radius $R_{200}$:
\begin{equation}
V_{200} = \left(\frac{GM_{200}}{R_{200}}\right)^{1/2}.
\label{eq:circular_velocity}
\end{equation}
For the NFW profile the ratio $V_{\rm max}/V_{200}$ is directly related to the halo concentration, $c_{200}$, by,
\begin{equation}
\frac{V_{\rm max}}{V_{200}} = \left(\frac{0.216c_{200}}{f(c_{200})}\right)^{1/2},
\label{eq:prada}
\end{equation}
where the function $f(c)$ is given by the following:
\begin{equation}
f(c) = \ln(1+c)-\frac{c}{(1+c)}.
\label{eq:c_function}
\end{equation}
Using only the measurements of $V_{\rm max}$ and $R_{200}$ (or $M_{200}$), Eqs.~(\ref{eq:circular_velocity})-(\ref{eq:c_function}) can be combined and solved numerically to estimate the concentration.

One challenge of this method that we encountered was that for some haloes in our sample $V_{200}>V_{\rm max}$. Our estimate of $V_{\rm max}$ includes only particles bound to the central, dominant subhalo of the FOF group, while $V_{200}$ has been calculated using all particles, bound or unbound, contained within $R_{200}$. Therefore it is possible that some large substructure found in the outer regions of the halo or some additional unbound particles flying past the halo could result in this inequality. We did not include P12 measurements of the concentration for haloes with $V_{200}>V_{\rm max}$.

Meanwhile the S08 method involves measuring the mean overdensity within $R_{\rm max}$ in units of the present-day critical density, $\rho_{\rm crit}$:
\begin{equation}
\delta_{\rm V} = \frac{\bar{\rho}(R_{\rm max})}{\rho_{\rm crit}} = 2\left(\frac{V_{\rm max}}{H_0R_{\rm max}}\right)^2.
\label{eq:rmax_density}
\end{equation}
The characteristic NFW overdensity, $\delta_{\rm c}$, is then calculated using,
\begin{equation}
\delta_{\rm c} = \frac{\rho_{\rm s}}{\rho_{\rm crit}} = 7.213\delta_{\rm V},
\label{eq:characteristic_overdensity}
\end{equation}
which is related to $c_{200}$ as follows:
\begin{equation}
\delta_{\rm c} = \frac{200}{3}\frac{c_{200}^3}{f(c_{200})}.
\label{eq:c_calculation}
\end{equation}
By starting with a measurement of $V_{\rm max}$ (or $R_{\rm max}$), the concentration can be estimated by combining Eqs.~(\ref{eq:rmax_density})-(\ref{eq:c_calculation}). Note that this method can be used for all haloes in our sample, including haloes with $V_{200}>V_{\rm max}$.

The main distinction between these two simplified methods is that the S08 method can give a more stable result which is independent of deviations from an NFW profile. This can be useful for simulations of a low resolution, since the S08 method is only sensitive to density changes around $R_{\rm max}$ and can therefore reproduce the density profile without too much noise. On the other hand, the P12 method uses information at $R_{200}$ as well as $R_{\rm max}$, so if the density profile deviates from an NFW profile outside $R_{\rm max}$ then this method can pick up this effect and yield a different result. So the P12 concentration estimate is expected to be more likely to agree with a measurement from a full fit of the NFW profile.

Because the density profile is slightly changed by $f(R)$ gravity for unscreened haloes, the differences between the methods will prove to be important for the results (see Sec.~\ref{sec:measurement_comparison}).

\subsubsection{Rescaled mass}
\label{sec:rescaled_mass}

A key aim of this work is to be able to predict how the halo concentration is affected in the screened and unscreened regimes of $f(R)$ gravity. The mass of the transition between these two regimes can be predicted using the $p_2$ parameter, as discussed in Sec.~\ref{sec:theory}. Therefore it is useful to rescale the mass by the transformation $\log_{10}(M_{500}M_{\odot}^{-1}h) \rightarrow \log_{10}(M_{500}M_{\odot}^{-1}h)-p_2\equiv\log_{10}(M_{500}/10^{p_2})$, such that negative values correspond to the unscreened regime and positive values correspond to the screened regime. It is essential to use a halo mass overdensity of 500 here, as discussed in Sec.~\ref{sec:simulations}. Note that $p_2$ can be measured using Eq.~(\ref{eq:p2}), together with Eqs.~(\ref{eq:R})-(\ref{eq:c1/c22}) which are used to evaluate the background scalar field.

\begin{figure*}
\centering
\includegraphics[width=1.0\textwidth]{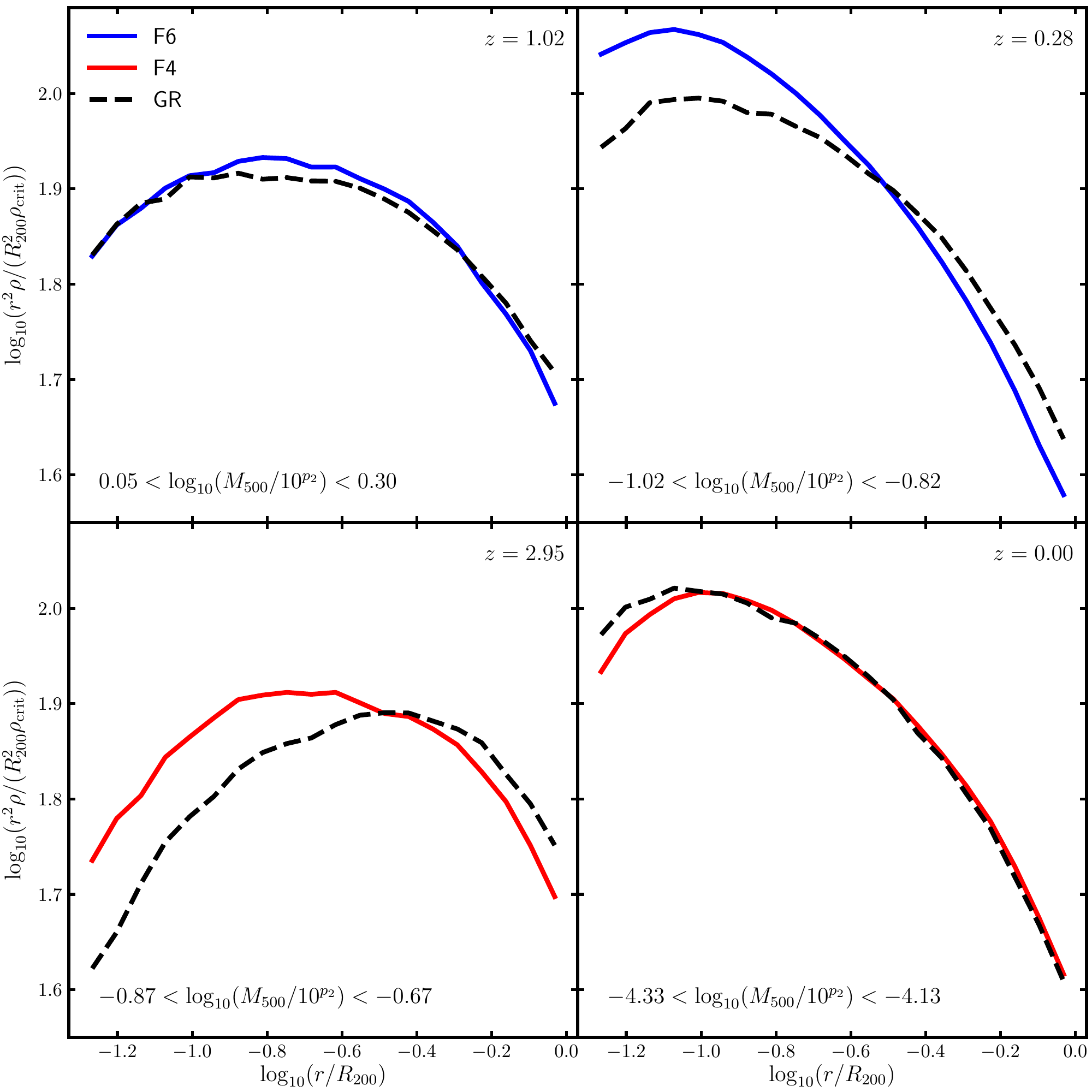}
\caption{Stacked density profile, scaled by $r^2$, measured using the median density profile of the haloes found within a particular mass bin. The data is from the \textsc{arepo} simulation (see Table \ref{table:simulations}), run for Hu-Sawicki $f(R)$ gravity with $n=1$ (\textit{solid line}) and GR (\textit{dashed line}). Each panel corresponds to a particular value of the present-day scalar field $f_{R0}$, where F6 and F4 correspond to $|f_{R0}|=10^{-6}$ and $|f_{R0}|=10^{-4}$, respectively, and a particular redshift $z$. The mass bins used cover the same range of halo masses, $M_{500}$, for the $f(R)$ and GR data, and each panel corresponds to a unique range of $\log_{10}(M_{500}/10^{p_2})$ values (annotated), where $p_2$ is evaluated with Eq.~(\ref{eq:p2}) using the values $|f_{R0}|$ and $z$ of that panel.}
\label{fig:stacked_profiles}
\end{figure*}

We motivate the rescaling approach in Fig.~\ref{fig:stacked_profiles}, which shows the stacked density profiles of the $f(R)$ haloes for four bins of $\log_{10}(M_{500}/10^{p_2})$, shown across four panels. The GR data for the same redshifts is also shown in each panel for a comparison. The density profile has been scaled by $r^{2}$ such that it peaks at distance $r=R_{\rm s}$ from the halo centre. This allows the concentration to be easily read off from the peak of the data, given that this is equal to $R_{200}/R_{\rm s}$.

The $f(R)$ data in the panel at the top-left corresponds to haloes that are partially screened. This is because the values of $\log_{10}(M_{500}/10^{p_2})$ used are just slightly above zero, and because the transition from screened to unscreened is smooth and gradual these haloes are therefore unscreened in the outer, less-dense regions but still screened in the inner, denser regions. This means that the particles falling towards the intermediate regions from the outer regions of the halo feel a stronger gravitational pull. However, the particle motions in the innermost regions of the halo are not affected by the fifth force, since this region is still screened. Therefore in $f(R)$ gravity the density at the outer regions is lower and the density at the intermediate regions is greater than in GR, but the density at the innermost regions is unaffected. This can actually lead to a larger scale radius $R_{\rm s}$ in $f(R)$ gravity than in GR, resulting in a lower value of the concentration. However the deviation between the $f(R)$ and GR density profiles in this regime of $\log_{10}(M_{500}/10^{p_2})$ is still quite small.

The top-right and bottom-left panels show regimes in which the entire halo has become unscreened. It is likely that these haloes have only recently entered the unscreened regime, since the values of $\log_{10}(M_{500}/10^{p_2})$ used are negative but still quite close to zero. Because the entire halo is now unscreened, the particles within both the inner and outer regions of the halo feel a stronger pull of gravity and thus fall towards the halo centre with a greater acceleration. Therefore the density is greater at the inner regions and lower at the outer regions than in GR. This results in a scale radius $R_{\rm s}$ that is smaller in $f(R)$ gravity, and so the concentration is greater than in GR. This regime of $\log_{10}(M_{500}/10^{p_2})$ has the greatest deviation between the $f(R)$ and GR density profiles.

Finally, the bottom-right panel shows data that is deep within the unscreened regime. This is because the values of $\log_{10}(M_{500}/10^{p_2})$ used are negative and much lower than the values spanned by the top-right and bottom-left panels. Therefore the haloes in this bin are likely to have been unscreened for a significant period of time. Interestingly, the density profiles in GR and $f(R)$ gravity are in reasonable agreement in all regions apart from the innermost region, in which the GR haloes are more dense. One possibility is that in $f(R)$ gravity the particles that initially fall into the halo centre gain a higher velocity during this substantial period of enhanced gravitational acceleration \citep{Shi:2015aya}, such that they are unable to settle into orbits at the innermost regions. As a result the scale radius is larger in $f(R)$ gravity than in GR, such that the concentration is greater in GR.

From these results, it is clear that the variable $\log_{10}(M_{500}/10^{p_2})$ can be a useful measure of the amount of screening of a halo, and so the model used to predict the enhancement of the concentration (Sec.~\ref{sec:general_model}) has been measured with respect to this variable. Plotting all results as a function of this variable also allows the combination of data with different values of $f_{R0}$ and $z$, since these are encapsulated by $p_2$ (see Sec.~\ref{sec:results}). 

\subsubsection{Concentration enhancement}
\label{sec:c_enhancement_measurement}

The GR data of each simulation has been outputted at snapshots with the same redshifts as the $f(R)$ data. Therefore the concentration enhancement of the haloes in a particular mass bin can be evaluated by first computing the median concentrations using the $f(R)$ and GR data from that bin, and then taking a ratio of these quantities. The choice of binning, the measurement of the median concentration and its error, and the evaluation of the concentration enhancement and its error is discussed in this section.

In order to evaluate the concentration enhancement for the haloes of a particular snapshot, the absolute concentrations of all haloes in $f(R)$ gravity and GR were measured using the methods discussed in Sec.~\ref{sec:c_measurement}. The haloes in each model were then binned by the halo mass $M_{500}$, with the binning chosen such that all bins are of equal width when viewed on a logarithmic axis apart from the highest-mass bin, which was allowed to be wide enough to contain at least 75 haloes in both $f(R)$ gravity and GR. For a given snapshot the same bins are used for both the GR and $f(R)$ gravity datasets, and the number of bins used is the greatest possible number such that each bin contains at least 75 haloes for both models.

The decision to use a wider highest-mass bin was taken due to the much smaller halo count at higher masses. The choice of having at least 75 haloes in each bin ensured that a balance was found between having enough bins with which to fit a reliable trend and keeping the errors low for each bin. This decision followed some tests which found that a minimum of 50 haloes per bin generates bins with a scatter that is too large, while a minimum of 100 haloes per bin results in very few bins, particularly in the \textsc{arepo} and Diamond data which do not contain as many haloes as the other simulations.

In each bin, the mean logarithm of the halo mass and the median concentration was measured for both $f(R)$ gravity and GR. The average of the two mean mass logarithms has been used to represent the mass of each bin. The error of the median concentration was evaluated using jackknife resampling, using the method discussed in \cite{jackknife} with 20 resamples. In this method the haloes of a snapshot are randomly split into 20 sub-volumes, then 20 resamples are created by systematically removing one sub-volume at a time. The haloes of each resample are split into the same set of mass bins as above, then the median concentration is measured for each bin of each resample. The one standard deviation error of the median concentration is then found by calculating the square root of the variance of the 20 values for each bin, which is multiplied by 19 to account for the lack of independence of the resamples. 


\begin{figure*}
\centering
\includegraphics[width=0.90\textwidth]{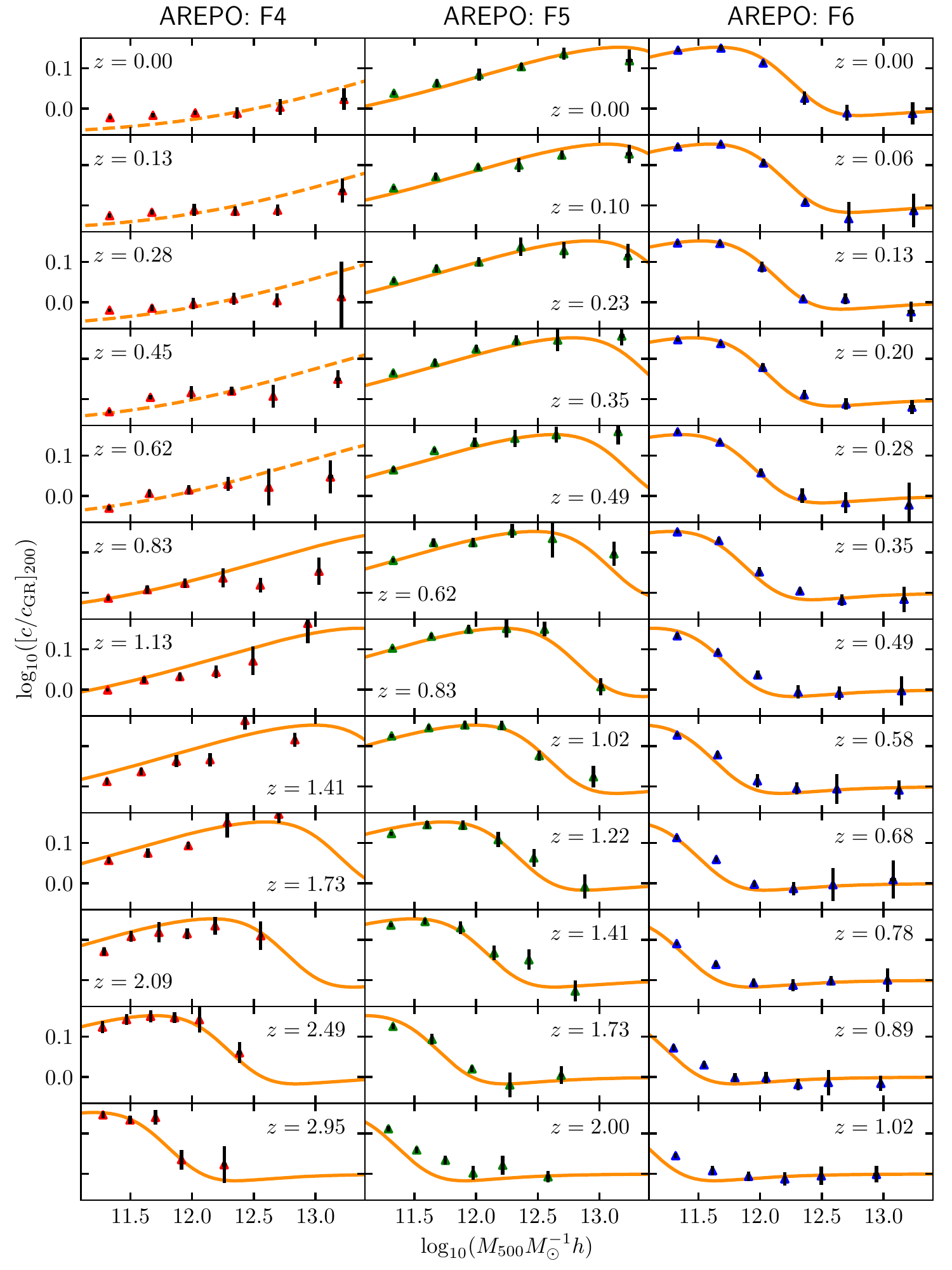}
\caption{Ratio of the median concentrations of $f(R)$ gravity and GR as a function of the halo mass, for Hu-Sawicki $f(R)$ gravity with $|f_{R0}|$ = $10^{-4}$ (F4, \textit{left column}), $10^{-5}$ (F5, \textit{middle column}) and $10^{-6}$ (F6, \textit{right column}) at various redshifts, $z$, as annotated. The parameter $p_2$ is evaluated using the values $|f_{R0}|$ and $z$ of each panel. Only haloes with mass $M_{500}>1.52\times10^{11}h^{-1}M_{\odot}$ from the modified \textsc{arepo} simulation (see Table \ref{table:simulations}), have been plotted. The one standard deviation error bars are shown. Predictions have been plotted (\textit{solid line}) for most snapshots, and are measured using the fit of Eq.~(\ref{eq:skewtanh}) to the data of Fig.~\ref{fig:skewtanh_fit}. The optimal parameter values from this fit are given in Table \ref{table:fitting}. In panels corresponding to data excluded from the fit, predictions are still shown, but in dashed lines.}
\label{fig:arepo_matrix}
\end{figure*}



\begin{figure*}
\centering
\includegraphics[width=0.90\textwidth]{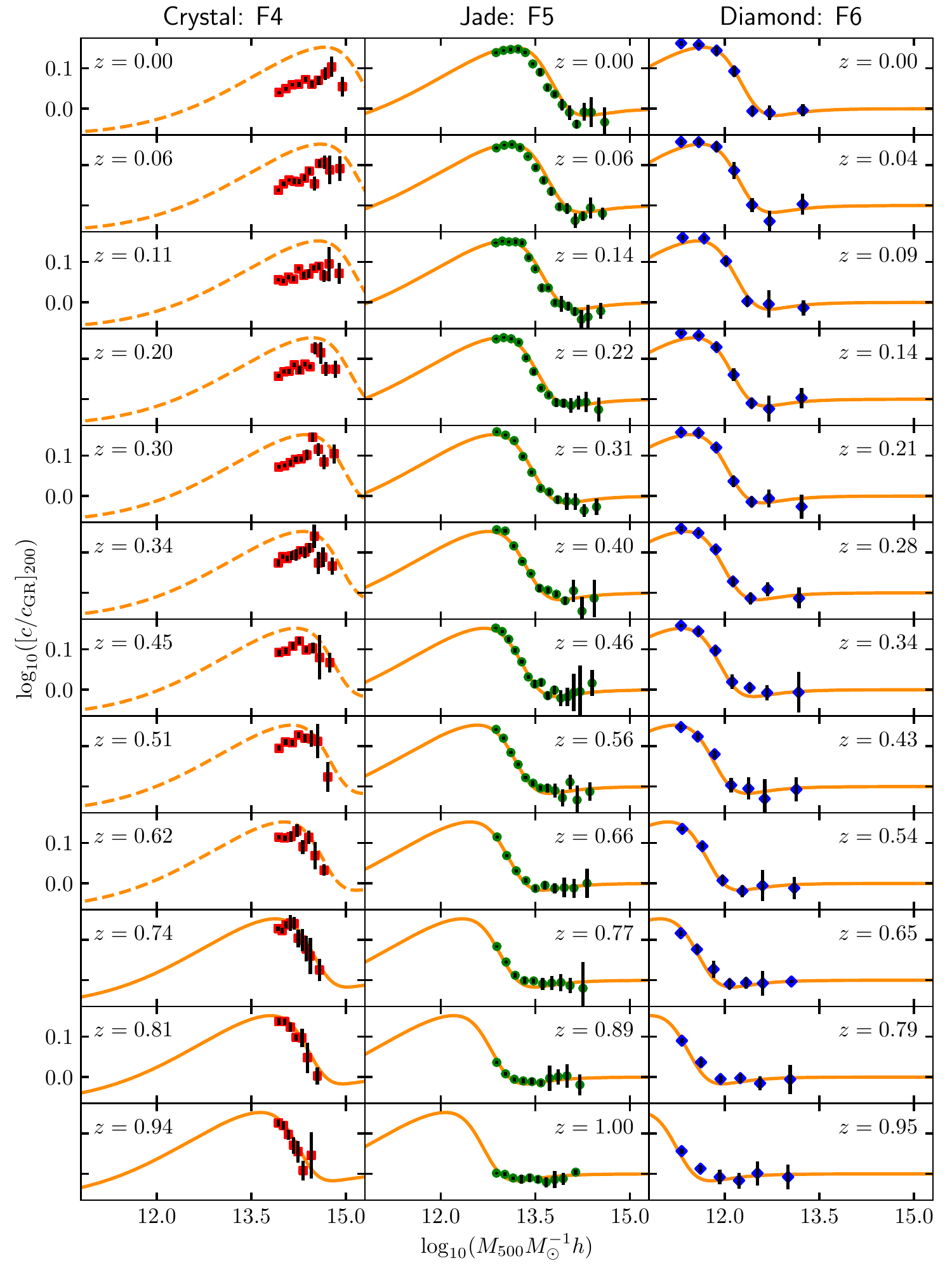}
\caption{Ratio of the median concentrations of $f(R)$ gravity and GR as a function of the halo mass, for Hu-Sawicki $f(R)$ gravity with $|f_{R0}|$ = $10^{-4}$ (F4, \textit{left column}), $10^{-5}$ (F5, \textit{middle column}) and $10^{-6}$ (F6, \textit{right column}) at various redshifts, $z$, as annotated. The parameter $p_2$ is evaluated using the values $|f_{R0}|$ and $z$ of each panel. Only haloes with mass $M_{500}>(7.78\times10^{13},6.64\times10^{12},1.52\times10^{11})h^{-1}M_{\odot}$ have been plotted for the Crystal (\textit{left column}), Jade (\textit{middle column}) and Diamond (\textit{right column}) modified \textsc{ecosmog} simulations respectively, the specifications of which are provided by Table \ref{table:simulations}. The one standard deviation error bars are shown. Predictions have been plotted (\textit{solid lines}) for most snapshots, and are measured using the fit of Eq.~(\ref{eq:skewtanh}) to the data of Fig.~\ref{fig:skewtanh_fit}. The optimal parameter values from this fit are given in Table \ref{table:fitting}. In panels corresponding to data excluded from the fit, predictions are still shown, but in dashed lines.}
\label{fig:ecosmog_matrix}
\end{figure*}


Finally, the logarithm of the ratio of the median concentration values in $f(R)$ gravity and GR, $\log_{10}\left(\left[c/c_{\rm GR}\right]_{200}\right)$, was evaluated to obtain the concentration enhancement. The error of the enhancement was measured by combining the median concentration errors in quadrature to find the error of the ratio, then the error of the logarithm of this ratio. Treating the errors as independent is justified here because the particle positions in collapsed structures are uncorrelated in these two simulations even though they start from the same initial conditions. This data is shown by the symbols in Figs.~\ref{fig:arepo_matrix} and \ref{fig:ecosmog_matrix} for an arbitrary selection of snapshots and models for each simulation. Note that the lines plotted in each panel correspond to predictions from our general model, which is discussed in Sec.~\ref{sec:general_model}. For each column the snapshots that are shown span the full range of available redshifts (see Table \ref{table:simulations}). For each bin the concentration enhancement and its error bar has been measured using the methods discussed above. The use of a wider highest-mass bin in each snapshot, as discussed above, can also be seen in each panel. Note that the same GR data is used when measuring the concentration enhancement for different $f(R)$ gravity models of the same simulation, although the binning scheme that is used may vary slightly.

\section{Results}
\label{sec:results}

The results shown in Figs.~\ref{fig:arepo_matrix} and \ref{fig:ecosmog_matrix} are plotted against the halo mass $M_{500}$. However, as discussed in Sec.~\ref{sec:rescaled_mass}, performing the transformation $\log_{10}(M_{500}M_{\odot}^{-1}h) \rightarrow \log_{10}(M_{500}/10^{p_2})$ converts the mass into a rescaled form in which negative values roughly correspond to the unscreened regime of halo mass and positive values roughly correspond to the screened regime. One advantage of this mass rescaling is that the $f(R)$ gravity model, redshift and cosmological parameters $\Omega_{\rm M}$ and $\Omega_{\Lambda}$ are all encapsulated by the $p_2$ parameter (Eq.~\ref{eq:p2}), so plotting against this rescaled mass can allow data from different $f(R)$ gravity models and even from simulations run for different cosmologies to be combined and plotted together in order to extract general trends. Note that there is a unique $p_2$ value for every combination of $z$, $f_{R0}$, $\Omega_{\rm M}$ and $\Omega_{\rm \Lambda}$, so, for example, all of the data in a particular panel of Fig.~\ref{fig:arepo_matrix} or \ref{fig:ecosmog_matrix} would be shifted by the same amount along the $\log_{10}(M_{500}M_{\odot}^{-1}h)$ axis when applying the transformation to the mass. 

We first used this plotting scheme to look into the three methods for measuring the concentration (see Sec.~\ref{sec:c_measurement}) to see how the choice of measurement can affect the results. This is discussed in Sec.~\ref{sec:measurement_comparison}. Then, focusing on the data produced from performing direct fitting of the NFW profile to individual haloes, a general model to describe the enhancement of the halo concentration in $f(R)$ gravity was sought and is discussed in Sec.~\ref{sec:general_model}.

\subsection{Concentration measurement comparison}
\label{sec:measurement_comparison}

\begin{figure*}
\centering
\includegraphics[width=1.0\textwidth]{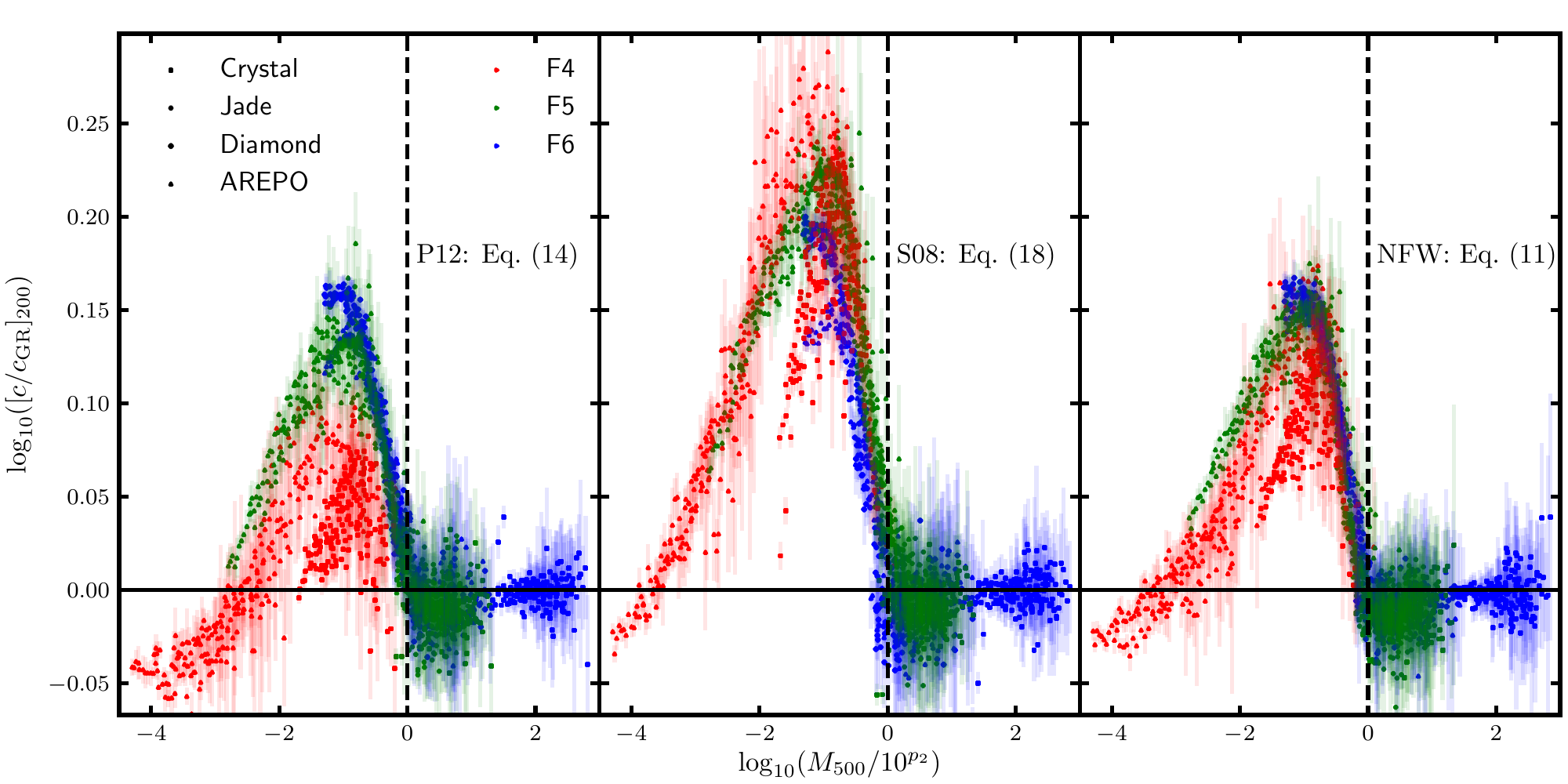}
\caption{Ratio of the median concentrations of $f(R)$ gravity and GR as a function of the rescaled mass, $\log_{10}(M_{500}/10^{p_2})$, for Hu-Sawicki $f(R)$ gravity with $|f_{R0}|$ = $10^{-4}$ (\textit{red}), $10^{-5}$ (\textit{green}) and $10^{-6}$ (\textit{blue}) and $n=1$. The plotted data is from the simulations summarised by Table \ref{table:simulations}, sifted so that only haloes with more than 1000 particles enclosed within $R_{500}$ are included. The concentration of each halo has been calculated using the methods discussed in Sec.~\ref{sec:c_measurement}, namely the methods discussed by P12 (\textit{left}) and S08 (\textit{middle}) and by performing a direct fitting of the NFW profile to the halo density profiles (\textit{right}). The one standard deviation error bars are shown.}
\label{fig:3_panel}
\end{figure*}

The concentration of each halo was measured using each of the three methods presented in Sec.~\ref{sec:c_measurement}. For each of these measures, the data was binned for every snapshot of each $f(R)$ model and the concentration enhancement and its error was measured for each bin using the method discussed in Sec.~\ref{sec:c_enhancement_measurement}. All of this data was plotted together against $\log_{10}(M_{500}/10^{p_2})$ to yield Fig.~\ref{fig:3_panel}, in which the three panels correspond to the three methods of measuring $c_{200}$.

The data from each panel follows a similar general trend. There is approximately zero enhancement of the concentration in the screened regime, then at lower mass (entering the unscreened regime) the enhancement rises to a peak at $\log_{10}(M_{500}/10^{p_2}) \approx -1$, before dropping to a negative enhancement at $\log_{10}(M_{500}/10^{p_2}) \lesssim -3$, where the GR concentration exceeds the $f(R)$ concentration. There is also a small dip in the concentration enhancement for $0 < \log_{10}(M_{500}/10^{p_2}) < 1$. The stacked profiles of Fig.~\ref{fig:stacked_profiles} and the discussion in Sec.~\ref{sec:rescaled_mass} can provide a physical interpretation of this behaviour.

The most accurate measurement of $c_{200}$ is by performing a direct fitting of the NFW profile to the halo profiles, so the panel on the right in Fig.~\ref{fig:3_panel} is expected to give the most reliable result. Here, the three $f(R)$ gravity models all show a similar behaviour and even peak at the same enhancement, which is approximately 0.15. Only the F4 data from the Crystal simulation shows any deviation from this behaviour, as it appears to have a lower concentration enhancement than the rest of the data at $\log_{10}(M_{500}/10^{p_2}) \approx -1$. However the good agreement of most of the data means that a general model can be fitted using a portion of the data, as discussed in Sec.~\ref{sec:general_model}.

The P12 data reaches the same maximum enhancement of approximately 0.15. However, the disparity between the three $f(R)$ models is now greater, with the models each peaking at a different enhancement. F6 has the highest peak enhancement and F4 has the lowest peak. Note that the P12 data has a reduced sample which excludes all haloes with $V_{200}<V_{\rm max}$ (see Sec.~\ref{sec:c_measurement}). The S08 data reaches a greater maximum enhancement of approximately 0.25, and shows an opposite trend in terms of the order of the models: F4 now has the highest peak enhancement and F6 has the lowest peak.

The difference in the results from the P12 and S08 methods is not surprising, yet it is significant. As shown in Fig.~\ref{fig:stacked_profiles}, data at $\log_{10}(M_{500}/10^{p_2}) \approx -1$ has a greater density at the inner regions of the halo and a lower density at the outer regions compared with GR, due to the enhanced acceleration of the in-falling particles. The S08 method only uses data from $R_{\rm max}$ to measure $c_{200}$, whereas the P12 method uses data from $R_{\rm max}$ and $R_{200}$. Being at a smaller distance from the halo centre, the mass enclosed by $R_{\rm max}$ is more affected by the in-fall of particles than the mass enclosed by $R_{200}$. So $V_{\rm max}$ is more affected than $V_{200}$, with the result that the S08 method measures a higher $f(R)$ concentration than the other methods. The P12 method is closer to actually fitting a density profile to the full extent of the halo, and so it yields a closer result to the full NFW fitting. 

These results indicate that the choice of method for measuring the concentration can be very important in modified gravity models. The three methods discussed in this work only agree perfectly for ideal NFW profiles, and therefore only the direct NFW fitting should be used for realistic cases. Note that this statement assumes that the halo density profiles in $f(R)$ gravity can still be well described by the NFW profile, which needs to be checked explicitly (see below). However, even if NFW is no longer valid, approximate methods such as P12 and S08 should not be used instead of full NFW fitting either as they are derivatives of the latter. Indeed, using the S08 method in $f(R)$ gravity could lead to a measurement of the halo concentration that is up to 26\% greater than from performing a fit. We therefore elected to use the direct NFW fitting method for all other results in this work. Note that in Sec.~\ref{sec:general_model} we include a check of the validity of the NFW profile in $f(R)$ gravity.

\subsection{General model for the concentration enhancement}
\label{sec:general_model}

\begin{figure*}
\centering
\includegraphics[width=1.0\textwidth]{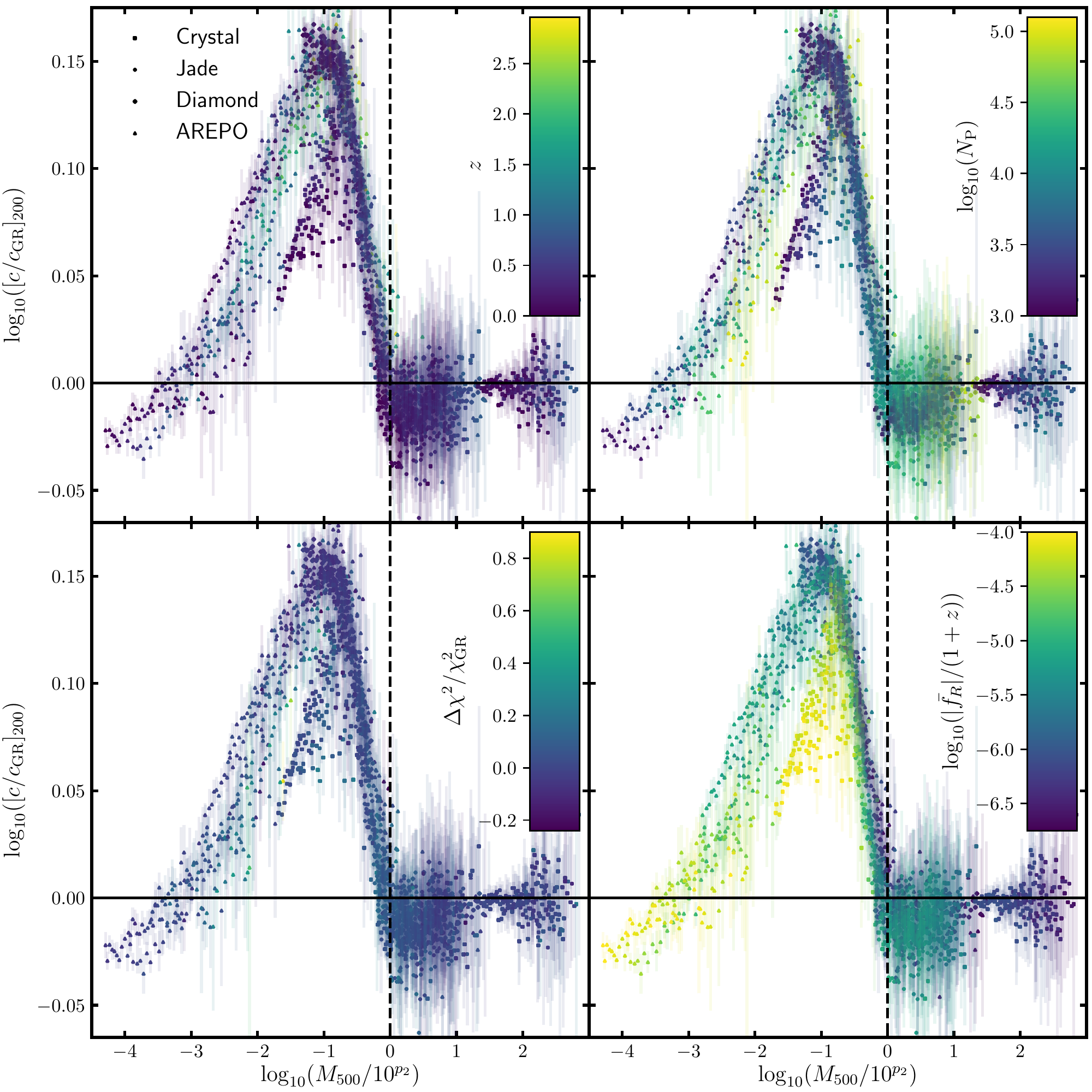}
\caption{Ratio of the median concentrations of $f(R)$ gravity and GR as a function of the rescaled mass, $\log_{10}(M_{500}/10^{p_2})$, for Hu-Sawicki $f(R)$ gravity with $|f_{R0}|$ = $10^{-4}$, $10^{-5}$ and $10^{-6}$ and $n=1$. The plotted data is from the simulations summarised by Table \ref{table:simulations}, sifted so that only haloes with more than 1000 particles enclosed within $R_{500}$ are included. The one standard deviation error bars are shown. The data is coloured by (\textit{clockwise from top left}) the redshift, the mean number of particles within $R_{500}$, the logarithm of a combination of the scalar field and redshift, $|\bar{f}_R|/(1+z)$, and the fractional difference of chi squared measures generated by NFW fits of the halo profiles.}
\label{fig:4_panel}
\end{figure*}

Before applying a fitting formula to the data it was useful to check whether certain factors could be affecting the shape of the trend, therefore the data was coloured via various schemes which are shown in Fig.~\ref{fig:4_panel}.

The top-left and top-right panels show the colourings by redshift and the particle number within each halo, respectively. These can both be viewed as tests of the effect of the halo resolution. For example, haloes with fewer particles can be prone to resolution effects at the innermost and outer regions, where the density can be underestimated. Haloes are less dense and more diffuse at higher redshift, which therefore leads to a greater exposure to these effects. As discussed in Sec.~\ref{sec:c_measurement}, in an effort to prevent these effects we restricted the fitting of the NFW profile to the radial range $0.05R_{200}$ to $R_{200}$. Furthermore we only include haloes that contain at least 1000 particles within $R_{500}$ in our sample. The coloured data of Fig.~\ref{fig:4_panel} suggests that these measures were sufficient, as it can be seen that even data at $z \approx 3$ agrees with the main trend and every part of the trend consists of haloes with both low and high particle numbers. Therefore even for the F4 Crystal data, low resolution is unlikely to be the reason for any disparity with the main trend.

It is significant that we are able to use redshifts up to $z=3$ for the \textsc{arepo} data. For a given $f_{R0}$ value, haloes at high $z$ are more screened than haloes at low $z$. This is because the magnitude of the background scalar field $\bar{f}_R$ grows as a function of time, such that haloes of a given mass will eventually go from being screened to unscreened. For $f(R)$ gravity models with a stronger scalar field (greater $|f_{R0}|$), haloes at a given redshift are more unscreened and therefore have a lower $\log_{10}(M_{500}/10^{p_2})$ value. However, models that are stronger than F4 are infeasible given current constraints on $f(R)$ gravity, and the minimum redshift that is available is $z=0$. Therefore the minimum value of $\log_{10}(M_{500}/10^{p_2})$ that can be plotted is only limited by the simulation resolution, as only haloes with lower mass can exist at lower values of this rescaled mass; similarly, the maximum value of $\log_{10}(M_{500}/10^{p_2})$ is limited by the box size. For each of the $f(R)$ gravity models tested in this analysis, the range of redshifts used provides a range of $\log_{10}(M_{500}/10^{p_2})$ that extends from the lowest value that is possible at halo mass $M_{500}=1.52\times10^{11}h^{-1}M_{\odot}$ to values in the screened regime, at which there is approximately no enhancement of the concentration compared with GR and so the concentration is much easier to predict. A weaker model of $f(R)$ gravity would likely exist close to or within the screened regime for $M_{500} \geq 1.52\times10^{11}h^{-1}M_{\odot}$ at $z=0$. Therefore, given that all three models tested in this work show excellent agreement for $-0.5 \leq \log_{10}(M_{500}/10^{p_2}) \leq 0.0$, it seems that a fit of this trend should be general for $M_{500} \geq 1.52\times10^{11}h^{-1}M_{\odot}$ for arbitrary values of $f_{R0}$ that are allowed by current constraints.

For every NFW fit we stored the $\chi^2$ value, which is measured by summing the squared residuals of the 20 radial bins. Storing the median $\chi^2$ value for every mass bin, the GR and $f(R)$ values of the latter were then combined to generate the fractional $\chi^2$ difference. The bottom-left panel of Fig.~\ref{fig:4_panel} shows the data coloured by this measure, and can therefore be seen as a test of the validity of the NFW profile in $f(R)$ gravity. The colouring shows that the goodness-of-fit of the NFW profile for most haloes in $f(R)$ gravity is within 20\% of the goodness-of-fit in GR. The colour-bar here shows the full range of fractional differences that were observed in the data, and we note that only a very small minority of haloes have a $\chi^2$ that is almost 90\% higher than in GR. These results are promising, and imply that systematics induced through the fitting of the NFW profile are unlikely to impact on the scatter of the halo concentration in $f(R)$ gravity.

Finally, the data was coloured by the logarithm of $|\bar{f_R}|/(1+z)$, and this is shown in the bottom-right panel of Fig.~\ref{fig:4_panel}. In \cite{Mitchell:2018qrg} it was found that complicated $f(R)$ gravity effects, including screening, can effectively be described by this useful parameter. It is therefore useful to see how the enhancement of the halo concentration at screened and unscreened regimes can depend on this. An interesting observation is that bins with $|\bar{f}_R|/(1+z) \lesssim 10^{-4.5}$ are in excellent agreement with a smooth trend for $-4 \lesssim \log_{10}(M_{500}/10^{p_2}) \lesssim 3$. The F6 data, which reaches a peak enhancement at $z \approx 0$, shows very good agreement with the F5 data, and both models agree well with the higher-$z$ F4 data. Therefore if a cut is made so that only data with $|\bar{f}_R|/(1+z) \leq 10^{-4.5}$ is used in the fitting, then, at least for halo masses $M_{500}\geq1.52\times10^{11}h^{-1}M_{\odot}$, a general model can be created that applies to arbitrary $f_{R0}$ values provided $|\bar{f}_R|/(1+z) \leq 10^{-4.5}$. For models with $|f_{R0}|>10^{-4.5}$ the concentration enhancement does not follow the same trend, and therefore cannot be described by the universal fitting formula below. However, we note that models with $|f_{R0}|>10^{-4.5}$ have already been strongly disfavoured by observations. 

\begin{figure*}
\centering
\includegraphics[width=0.7\textwidth]{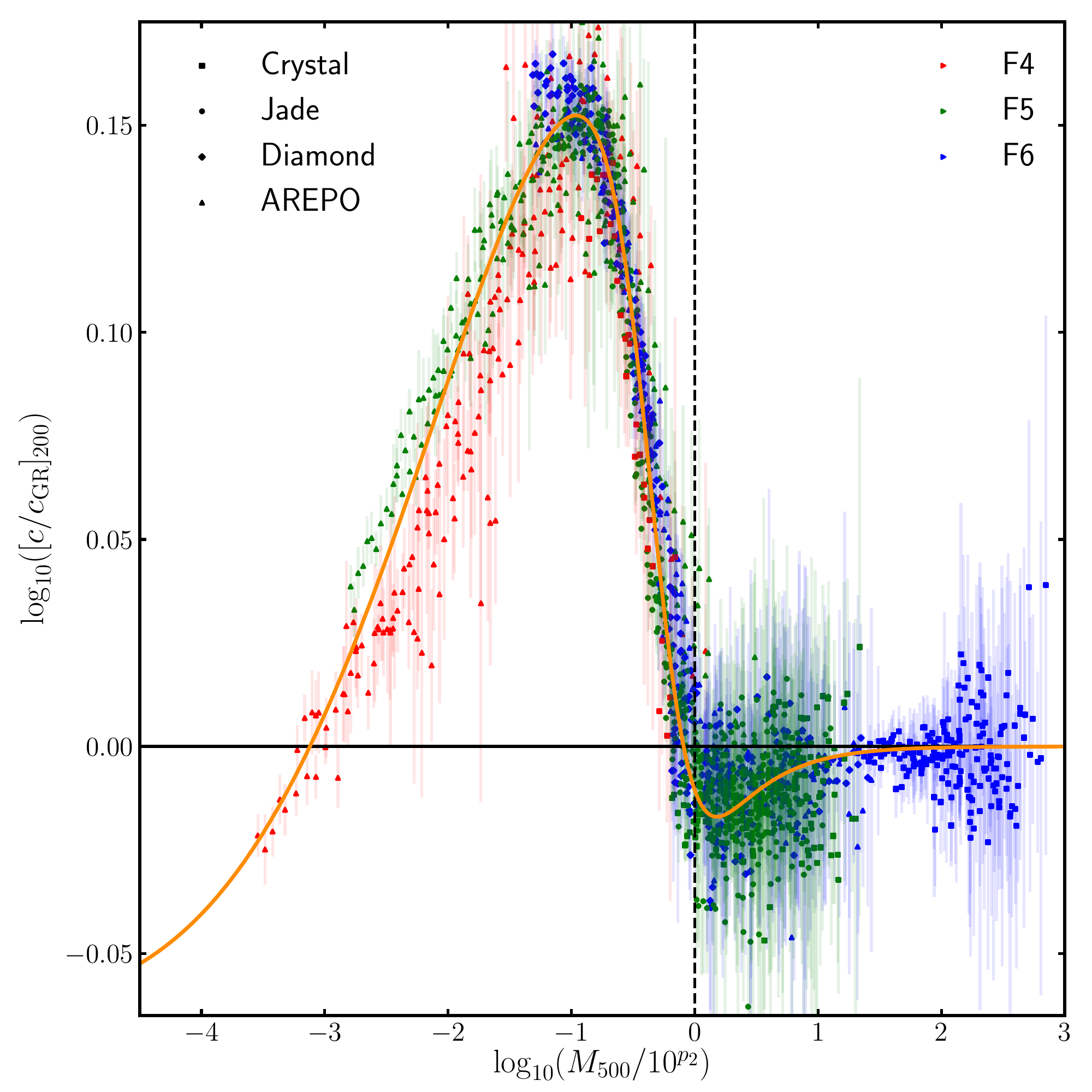}
\caption{Ratio of the median concentrations of $f(R)$ gravity and GR as a function of the rescaled mass, $\log_{10}(M_{500}/10^{p_2})$, for Hu-Sawicki $f(R)$ gravity with $|f_{R0}|$ = $10^{-4}$ (\textit{red}), $10^{-5}$ (\textit{green}) and $10^{-6}$ (\textit{blue}) and $n=1$. Only data with $|\bar{f}_R|/(1+z) \leq 10^{-4.5}$ has been included and fitted with Eq.~(\ref{eq:skewtanh}). This fit is shown by the trend-line, and the optimised parameter values are listed in Table \ref{table:fitting}. The plotted data is from the simulations summarised by Table \ref{table:simulations}, sifted so that all haloes have at least 1000 particles enclosed within $R_{500}$. The one standard deviation error bars are shown.}
\label{fig:skewtanh_fit}
\end{figure*}

Cleaning the data to remove bins with $|\bar{f}_R|/(1+z) > 10^{-4.5}$ and re-plotting yields Fig.~\ref{fig:skewtanh_fit}. This shows a clear trend. As $\log_{10}(M_{500}/10^{p_2})$ is reduced from value 3, the concentration enhancement, which is initially zero, drops slightly to a negative enhancement. Continuing into the unscreened regime the enhancement then rises to a distinct peak with value $\approx0.15$ at $\log_{10}(M_{500}/10^{p_2})\approx-1$, before dropping down to negative values again for $\log_{10}(M_{500}/10^{p_2})\lesssim-3$. From the discussion in Sec.~\ref{sec:rescaled_mass} and the results in Fig.~\ref{fig:stacked_profiles}, the above behaviour is physical and should therefore be fully included in the fitted model. 

In selecting a suitable fitting formula, both the screened and unscreened regimes of the data were considered. The data in the unscreened part of Fig.~\ref{fig:skewtanh_fit} shows good agreement with a skewed Gaussian curve, whose steepness is different on the two sides of the peak. This requires five parameters: a scaling $\lambda$ of the height of the curve, a shift $\gamma$ along the $\log_{10}\left(\left[c/c_{\rm GR}\right]_{200}\right)$ axis, a width $\omega_{\rm s}$, a skewness parameter $\alpha$ and a parameter $\xi_{\rm s}$ to describe the location with respect to the $\log_{10}(M_{500}/10^{p_2})$ axis. As discussed in Sec.~\ref{sec:rescaled_mass} and from examining the top-left panel of Fig.~\ref{fig:stacked_profiles}, there is a physical motivation that the concentration should dip slightly at halo masses just greater than $10^{p_2}h^{-1}M_{\odot}$. Therefore the model would have to include a minimum in this regime. This was achieved by multiplying the skewed Gaussian with a $\tanh$ curve, which takes value 1 at low values of $\log_{10}(M_{500}/10^{p_2})$ and drops to value 0 at high values. This induces a further two parameters: a location $\xi_{\rm t}$ and a width $\omega_{\rm t}$ with respect to the $\log_{10}(M_{500}/10^{p_2})$ axis. The $\tanh$ curve also ensures that the model tends to zero at higher $\log_{10}(M_{500}/10^{p_2})$. This model would gradually level out at $\log_{10}(M_{500}/10^{p_2}) < -4$, however this is not necessarily how the concentration enhancement would behave in this regime. The only way to understand this would be to run higher-resolution simulations so that lower-mass haloes can be investigated.

By taking the above considerations into account, we arrive at a 7-parameter fitting formula, which is given by,
\begin{equation}
y(x) = \frac{1}{2}\left(\frac{\lambda}{\omega_{\rm s}}\phi(x')\left[1+\rm{erf}\left(\frac{\alpha x'}{\sqrt[]{2}}\right)\right]+\gamma\right)\left(1-\tanh\left(\omega_{\rm t}\left[x+\xi_{\rm t}\right]\right)\right),
\label{eq:skewtanh}
\end{equation}
where $y=\log_{10}\left(\left[c/c_{\rm GR}\right]_{200}\right)$, $x'=(x-\xi_{\rm s})/\omega_{\rm s}$ and $x=\log_{10}(M_{500}/10^{p_2})$. The left-hand bracket of Eq.~(\ref{eq:skewtanh}) represents the skewed Gaussian curve, where $\phi(x)$ is the normal distribution:
\begin{equation}
\phi(x) = \frac{1}{\sqrt[]{2\pi}}\exp\left(-\frac{x^2}{2}\right).
\label{eq:normal_dist}
\end{equation}
This also includes a multiplication with the error function ${\rm erf}(x')$ in order to generate a skewed curve. The error function is given by,
\begin{equation}
{\rm erf}(x) = \frac{2}{\sqrt[]{\pi}}\int_0^xe^{-t^2}{\rm d}t.
\end{equation}
The fit of Eq.~(\ref{eq:skewtanh}) to the data of Fig.~\ref{fig:skewtanh_fit} was carried out by minimising the sum of the squared normalised residuals of the data points, where the residuals have been normalised by the one standard deviation error bars shown. The sum is evaluated in a way that treats different parts of the $\log_{10}(M_{500}/10^{p_2})$ range equally. This has been achieved by splitting this range into 13 equal-width bins. The squared normalised residuals are then weighted by the reciprocal of the number of data points in the current bin, prior to minimising the sum through varying the parameters. The optimal parameters are listed in Table \ref{table:fitting} and the corresponding fit is shown in Fig.~\ref{fig:skewtanh_fit}.

\begin{table*}
\centering

\small
\begin{tabular}{ ccccccc } 
 \toprule
 
 $\lambda$ & $\xi_{\rm s}$ & $\omega_{\rm s}$ & $\alpha$ & $\gamma$ & $\omega_{\rm t}$ & $\xi_{\rm t}$ \\

 \midrule

 $0.55\pm0.18$ & $-0.27\pm0.09$ & $1.7\pm0.4$ & $-6.5\pm2.4$ & $-0.07\pm0.04$ & $1.3\pm1.0$ & $0.1\pm0.3$ \\ 
 
 \bottomrule
 
\end{tabular}

\caption{Optimal parameter values and errors from the fit of Eq.~(\ref{eq:skewtanh}) to the data of Fig.~\ref{fig:skewtanh_fit}. The fit is carried out by first splitting the range of $\log_{10}(M_{500}/10^{p_2})$ into 13 equal-width bins. The squared normalised residuals of the data points are then weighted by the reciprocal of the number of data points in the current bin. The sum of these is minimised by varying the parameters.}
\label{table:fitting}

\end{table*}

We have also considered a weighted least squares fit which neglects the weighting of the squared normalised residuals described above. This results in a model that produces nearly identical predictions to the model shown in Fig.~\ref{fig:skewtanh_fit}. However, neglecting the weighting of the squared normalised residuals disfavours parts of the $\log_{10}(M_{500}/10^{p_2})$ range that contain fewer data points, including the \textsc{arepo} F4 data at $\log_{10}(M_{500}/10^{p_2})\lesssim-3$. Therefore, we only include results from the fitting described above.

In order to test our model, its predictions were generated for the data shown in Figs.~\ref{fig:arepo_matrix} and \ref{fig:ecosmog_matrix}. The predictions are shown by the plotted lines. Solid lines are used in snapshots which were used to generate the fit in Fig.~\ref{fig:skewtanh_fit} and dashed lines are used in snapshots excluded from the fit (snapshots with $|\bar{f}_R|/(1+z) > 10^{-4.5}$). Agreement is generally excellent between the data and the predictions in both figures. Agreement is reasonable even for the low-$z$ F4 snapshots of \textsc{arepo} that were not used to generate the model, as can be seen in Fig.~\ref{fig:arepo_matrix}. Some small disparity exists in the higher-$z$ F5 and F6 snapshots, where the data does not appear to agree with the predicted minimum in the screened regime. Again, there are probably some physical effects that result in subtly different trends at different redshifts. However, given the complexity of the behaviour of the halo concentration in chameleon $f(R)$ gravity and the simplicity of our modelling, the amount of agreement shown in these figures is indeed very good.

\section{Summary, Discussion and Conclusions}
\label{sec:conclusions}

The global properties of galaxy clusters are sensitive to cosmological parameters and gravitational models. For example, the cluster abundance depends largely on the growth rate of structure, which in turn is sensitive to the behaviour of the force of gravity over large scales. Galaxy clusters are therefore a powerful probe of modified gravity theories on cosmological scales, and the results of the many ongoing and upcoming high-impact galaxy and cluster surveys are set to revolutionise the tightness of these constraints. However the model-dependent theoretical predictions should be in a form that they can be safely confronted with the observational data without inducing bias and systematic effects in the final constraints. This is not so straightforward in tests of modified gravity theories, where the properties of galaxy clusters can be affected such that various $\Lambda$CDM results for the cluster observable-mass relations are no longer valid.

This paper is the second in a series of papers aiming to create a general framework which incorporates the various effects of modified gravity on the properties of galaxy clusters so that model parameters can be constrained without bias. In the first paper \citep{Mitchell:2018qrg} a general model was calibrated for the enhancement of the dynamical mass of clusters versus their true mass in HS $f(R)$ gravity with $n=1$. This was probed vigorously over a continuous range of scalar field values for models spanning $-6.5 \leq \log_{10}(|f_{R0}|) \leq -4.0$ using simulation data running up to $z=1$ and covering a wide range of halo masses. Perhaps most significantly, this model depends on just a single parameter, which is a combination of the background scalar field and redshift: $|\bar{f}_R|/(1+z)$. That work also introduced a parameter denoted $p_2$, which is given by Eq.~(\ref{eq:p2}), that can be used to predict the mass, $M_{500}=10^{p_2}h^{-1}M_{\odot}$, above which haloes are screened and below which haloes are unscreened. Note that this parameter encapsulates the values of $f_{R0}$, $z$, $\Omega_{\rm M}$ and $\Omega_{\Lambda}$, so that it encodes the cosmology dependence of chameleon screening, as well as the dependence on the $f(R)$ gravity parameter. It is a universal description that allows $f(R)$ models with different $f_{R0}$ to be studied in a unified way.

In this work, a model has been developed for the enhancement of the halo concentration in HS $f(R)$ gravity with $n=1$ using a suite of simulations that are summarised by Table~\ref{table:simulations}. The model is shown in Fig.~\ref{fig:skewtanh_fit}, and is given by Eq.~(\ref{eq:skewtanh}) with the parameter values listed in Table \ref{table:fitting}. It has been defined in terms of a useful rescaling of the halo mass, $M_{500}/10^{p_2}$, such that data from three different $f(R)$ gravity models again satisfy a universal description. These models have $\log_{10}(|f_{R0}|)=(-4,-5,-6)$, and the fitting was carried out using data from all simulation snapshots with $\log_{10}\left(|\bar{f}_R|/(1+z)\right)\leq-4.5$. This universal description is shown to have very good agreement with simulations for $M_{500}/10^{p_2}$ covering nearly 7 orders of magnitude, and covering five decades of the halo mass.

Our model has been tested by comparing its predictions of the enhancement of the concentration with an arbitrarily chosen set of snapshots from our simulations, as shown by the lines plotted in Figs.~\ref{fig:arepo_matrix} and \ref{fig:ecosmog_matrix}. These predictions show excellent agreement with the data for all snapshots, apart from the Crystal snapshots with $\log_{10}\left(|\bar{f}_R|/(1+z)\right)>-4.5$. This is not surprising given that this data was not used in the fit of the model. Having a general model that works for $\log_{10}\left(|\bar{f}_R|/(1+z)\right)\leq-4.5$ will prove very useful, particularly given that an analytical theoretical modelling was not available.

The data of Fig.~\ref{fig:skewtanh_fit} shows that  in the unscreened regime the enhancement of the concentration reaches a distinct peak as a function of the halo mass, but drops to negative values at lower mass, where the $f(R)$ concentration is less than the GR concentration. As shown by Fig.~\ref{fig:stacked_profiles}, such negative enhancement occurs because the innermost regions of the haloes are less dense in $f(R)$ gravity than in GR. This could be caused by the velocity gained by particles in haloes, which makes it difficult for them to settle into orbits at the central regions of the halo. Meanwhile in the screened regime of Fig.~\ref{fig:skewtanh_fit} there is a small dip in the concentration. Fig.~\ref{fig:stacked_profiles} suggests that this is caused by the halo being only partially screened, so that outer particles are moved further towards the centre of the halo while the inner regions remain screened. The density profile is therefore unaffected at the innermost regions but is greater at intermediate radii. Therefore the scale radius becomes greater, and fitting an NFW profile would then result in an estimate for the concentration that is lower in $f(R)$ gravity than in GR. All of these effects are incorporated by the fitted model of Eq.~(\ref{eq:skewtanh}).

Some further investigations were carried out which can be useful for further studies of the concentration in $f(R)$ gravity, and in other similar modified gravity theories. Firstly, in addition to applying a direct NFW profile fitting to each of the haloes to measure the concentration, two simplified approaches were also used, namely the methods that are used by \cite{Prada:2011jf} and \cite{Springel:2008cc}. The resulting enhancement of the concentration from using these two methods (shown in Fig.~\ref{fig:3_panel}) shows a difference from direct NFW fitting. This is due to the effects of $f(R)$ gravity on the internal density profile, which means that the choice of regions of the halo to use in measuring the concentration becomes important. The method used by \cite{Springel:2008cc} only requires the mass enclosed by the orbital radius with the maximum circular velocity. Being found at the inner regions of a halo, which become more dense as the halo becomes unscreened, this results in the concentration being overestimated by up to 26\%. From this, we conclude that only the direct NFW fitting should be used in $f(R)$ studies. Secondly, we looked at the validity of the NFW profile fitting in $f(R)$ gravity and found that, as shown by the bottom-left panel of Fig.~\ref{fig:4_panel}, for most haloes the $\chi^2$ measure for the fit is within 20\% of the GR measure, and for some haloes the fit is even better. Therefore the systematic effects caused by fitting the NFW profile in $f(R)$ gravity are unlikely to have a significant effect on the scatter of the concentration measure.

This work shows that the $p_2$ parameter defined by \cite{Mitchell:2018qrg} can indeed be very useful in the description and modelling of complicated effects in $f(R)$ gravity. In addition to its relatively simple one-parameter definition it may also allow the combining of data generated by simulations run for different cosmological parameters, as $p_2$ encapsulates the values of $\Omega_{\rm M}$ and $\Omega_{\Lambda}$. Indeed, the data for the concentration enhancement from \textsc{arepo} and Diamond F6 shows excellent agreement (see Fig.~\ref{fig:skewtanh_fit}), even though these two simulations were run for different cosmological parameters and using very different codes. It will be interesting to see where else $p_2$ can be used in $f(R)$ studies. Of particular interest would be to see how it can simplify the modelling of the HMF. The enhancement of the HMF in $f(R)$ gravity peaks at a particular halo mass which depends on the strength of the scalar field. A stronger scalar field allows higher-mass haloes to be unscreened, and therefore results in an enhancement of the HMF at a higher mass. At the very least, the mass of the peak enhancement of the HMF can be expected to be strongly correlated to $p_2$. The enhancement of the matter power spectrum could also be investigated via a similar treatment. We are currently working on these and will report the results in forthcoming works.

This work used data from four different simulations, allowing a wide range of resolutions to be used. However, one potential drawback is that these are run for dark matter only. It would therefore be useful to test these results using cluster data from full-physics hydrodynamical simulations run for $f(R)$ gravity. The requirement of these simulations has already been included in the general framework, as shown in Fig.~\ref{fig:flow_chart}, where full-physics hydrodynamical simulations will be used to investigate various observable-mass scaling relations in $f(R)$ gravity.

\section*{Acknowledgements}

We thank Matteo Cataneo for helpful discussions and comments. MAM is supported by a PhD Studentship with the Durham Centre for Doctoral Training in Data Intensive Science, funded by the UK Science and Technology Facilities Council (STFC, ST/P006744/1) and Durham University. CA, JH and BL are supported by the European Research Council via grant ERC-StG-716532-PUNCA. BL is additionally supported by STFC Consolidated Grants ST/P000541/1, ST/L00075X/1. This work used the DiRAC Data Centric system at Durham University, operated by the Institute for Computational Cosmology on behalf of the STFC DiRAC HPC Facility (\url{www.dirac.ac.uk}). This equipment was funded by BIS National E-infrastructure capital grant ST/K00042X/1, STFC capital grants ST/H008519/1, ST/K00087X/1, STFC DiRAC Operations grant ST/K003267/1 and Durham University. DiRAC is part of the National E-Infrastructure.




\bibliographystyle{mnras}
\bibliography{references} 





\bsp	
\label{lastpage}
\end{document}